\documentclass[letterpaper]{article}
\usepackage[margin=1.5in]{geometry}
\usepackage{natbib}
 \bibpunct[, ]{(}{)}{,}{a}{}{,}%
 \def\bibfont{\small}%
\usepackage{times}
\usepackage{amsfonts}
\usepackage{amsmath}
\usepackage{amssymb}
\usepackage{amsthm}
\usepackage{graphicx}
\usepackage{algorithm}
\usepackage{algpseudocode}
\usepackage{url}
\usepackage{multirow}
\usepackage{booktabs}
\usepackage{setspace,cellspace,booktabs}
\usepackage{subeqnarray}
\usepackage{mathdots}
\usepackage{enumitem}
\usepackage{mathptmx}
\usepackage{wrapfig}
\usepackage{color,soul}
\usepackage{bm}
\usepackage{longtable}
\usepackage{multicol}
\usepackage{appendix}
\usepackage{longtable}

\newtheorem{theorem}{Theorem}
\newtheorem{remark}{Remark}

\newcommand{\Rm}{\mathrm}

\newcommand{\Cal}{\mathcal}

\mathchardef\mhyphen="2D
\renewcommand{\v}[1]{\ensuremath{\mathbf{#1}}}

\title{A Nested Cross Decomposition Algorithm for Power System Capacity Expansion with Multiscale Uncertainties}

\author{Zhouchun Huang \thanks{College of Economics and Management, Nanjing University of Aeronautics and Astronautics, Nanjing, China.} 
\and Qipeng P. Zheng  \thanks{Department of Industrial Engineering and Management Systems, University of Central Florida, Orlando, FL, USA}
\and Andrew L. Liu \thanks{School of Industrial Engineering, Purdue University, West Lafayette, IN, USA.}}

\date{March 27, 2021}

\begin{document}

\maketitle

\begin{abstract}
  Modern electric power systems have witnessed rapidly increasing penetration of renewable energy, storage, electrical vehicles and various demand response resources. The electric infrastructure planning is thus facing more challenges due to the variability and uncertainties arising from the diverse new resources. This study aims to develop a multistage and multiscale stochastic mixed integer programming (MM-SMIP) model to capture both the coarse-temporal-scale uncertainties, such as investment cost and long-run demand stochasticity, and fine-temporal-scale uncertainties, such as hourly renewable energy output and electricity demand uncertainties, for the power system capacity expansion problem.  
  To be applied to a real power system, the resulting model will lead to extremely large-scale mixed integer programming problems, which suffer not only the well-known curse of dimensionality, but also computational difficulties with a vast number of integer variables at each stage. In addressing such challenges associated with the MM-SMIP model, we propose a nested cross decomposition algorithm that consists of two layers of decomposition, that is, the Dantzig-Wolfe decomposition and L-shaped decomposition. The algorithm exhibits promising computational performance under our numerical study, and is especially amenable to parallel computing, which will also be demonstrated through the computational results.
\end{abstract}

%\KEYWORDS{Electrical Power System, Expansion Planning, Multiscale, Multistage Stochastic Mixed Integer Programming, Nested Cross Decomposition, Parallel Computing}

\section{Introduction}
\label{cep:intro}
Global electricity consumption is increasing rapidly because of economic growth and rising population. 
It is forecasted by the U.S. Energy Information Administration that the electricity use will grow by nearly 80\% between 2018 and 2050 \citep{EIA2019}. This makes it necessary to expand generation capacity so as to meet future electricity demand. 
Meanwhile, various renewable energy sources, including solar, wind and hydroelectric, and new clean technologies have emerged in the past decade to reduce air pollutant emissions and to promote sustainability. Environmental regulations have led the electric industry to retire aging fossil fuel generators and to build more and more power plants with these new technologies \citep{HUANG20201036}. 
Along with more installed generators, the expansion of transmission capacity might be necessary to mitigate power system congestion. Especially for wind farms which are usually located in sparsely-populated areas, new transmission lines are needed to connect the generation resources and demand centers. Furthermore, energy storage technologies can be installed and used as a buffer between generation and consumption, thereby mitigating the unpredictability and intermittency associated with renewable sources.

The capacity expansion of power system infrastrues is a capital-intensive and long-lasting process. 
However, few existing models truly integrate the planning of generation, transmission and energy storage over a long planning horizon due to computational complexities, although the coordinated planning may produce more cost effective results than a separate or sequential decision process \citep{CoOpt16}. 
This motivates our work that co-optimizes all of the infrastructures and can be applied to large-scale systems through algorithm advances and high-performance computing.
In addition to its aging infrastructures, the electricity sector in a power system is usually facing a rising amount of volatility in different time scales. 
In the coarse temporal scale, the capital costs for new installations and long-run electricity demand growth rates are both uncertain and can only be estimated. In the fine temporal scale, the intermittent nature of renewable energy poses considerable challenges to system operation
as electricity supply and demand must be balanced at all times to ensure system reliability; and on the demand side, high penetration of plug-in electric vehicles and other demand response resources make future demand more unpredictable. These uncertainties need to be properly accounted for through the capacity expansion process, as otherwise the decisions might be either not cost minimizing or not sufficiently reliable to meet future demand \citep{LARA20181037, LiSiCo2018}.
The need for a model with a multiscale structure that can produce a uniform infrastructure expansion schedule to accommodate a wide range of possibilities in the future provides another motivation for our study.

This paper proposes a multistage and multiscale stochastic mixed integer programming (MM-SMIP) model for the power system capacity expansion problem. The model is multistage in terms that it makes capacity expansion decisions in different stages of time; and at a given stage, with all the past uncertainties realized, it seeks the best planning decisions subject to future uncertainties of investment costs and electricity demand. It is multiscale in the sense that it integrates detalied short-term unit commitment modeling through the planning process with the consideration of uncertainties in the fine temporal scale, such as renewable energy output intermittency and hour-to-hour electricity demand volatility, in addition to those in the coarse temporal scale. 
The resulting model is an extremely large-scale stochastic mixed-integer program, easily of multi-million variables and constraints when applied to a real-world system.
The multistage and multiscale features of the model, however,  make it highly structured and decomposable. We propose a nested cross decomposition (NCD) algorithm to fully exploit such features, which is also amenable to parallel computing.
There are two layers of decomposition for this problem. At the capacity expansion level, column generation is used to decompose the multistage expansion problem and iteratively generates feasible expansion plans.
The second layer of L-shaped decomposition is to separate the stochastic unit commitment problems from the scenario tree of the expansion planning through cutting plane approximation. Even though we can decompose the huge MM-SMIP into much smaller problems, there are still a vast number of them. We employ parallel computing here as the decomposed problems can be solved independently.

The remainder of this paper is organized as follows. A literature review is provided in Section \ref{sec:lit}. Section \ref{sec:mdl} introduces the structure and mathematical formulation of the MM-SMIP model. Section \ref{cep:sec:alg} discusses the NCD algorithm for the proposed model. Numerical studies on the model and parallel computing implementation of the algorithm are presented in Section \ref{sec:exp}. Section \ref{sec:con} concludes the papers and discusses future research.

\section{Literature Review}\label{sec:lit}
A planning process of capacity expansion is to identify infrastructure needs for a system to serve growing demand in the future.
There is an extensive literature on capacity expansion and long-term planning in general
\citep{Lu82, DaDeSeVe87, BeHiSm92, LiTi94, RaSiMo98, AhKiPa03,SiPhWo09} and specific applications, such as communication networks \citep{Laguna1998,Riis2002}, supply chain management \citep{SANTOSO200596}, project management \citep{Conejo2021}, manufacturing industries \citep{Gary1989,Sahinidis1992} and service industries \citep{Herman1994,Jena2017}.  

Planning for capacity expansion in an electric power system differs from the other applications, mainly because of various uncertainties from diverse resources %such as intermittent renewable energy generation. 
and many requirements and constraints (e.g., ramping and minimum-up and down time constraints) for securing power syste operations.
Stochastic optimization has been applied by a number of works as the framework for modeling electric utilities planning problems that involve uncertainty \citep{ShBi03,WaFl03,Giraldo2014,ZouJikai2018,Saeed2019}. 
However, none of them considers detailed operational features, which however become more and more important especially due to the increasing share of the renewable energy sources, along with the reliability requirement to balance electricity supply and demand at all times. In some other stochastic programming approaches \citep{SiPhWo09,Munoz2015,ChenLv2018}, the multiscale feature of uncertainties in the decision process has not been fully addressed. In fact, the study in \cite{PONCELET2016631} shows that, for the power system planning with a high penetration of intermittent renewable energy sources, the gains obtained by improving the temporal representation are even more than that by incorporating detailed operational constraints. %Both these simplifications have been shown to have a significant impact on the results.

There are a few studies that integrate operational modeling through the expansion planning process with consideration of detailed multiscale representation of uncertainties.
\cite{XiHoLiPeRe11} develop a multistage multiscale stochastic model considering simultaneously long-term capacity expansion and short-term dispatch decisions. Their approach considers the integration of wind farms only, and the benefit of considering multiscale uncertainties is verified by a case study on a small 3-bus system. 
\cite{PoGeSiScLaSt10} employ the multiscale modeling through an approximate dynamic programming approach to integrate the investment decisions and electric dispatch. Their model does not represent individual energy generators and is based on stochastic control theory, in which the demands are assumed to be simple stochastic processes, to render analytical tractability. 
\cite{LARA20181037} propose a multiscale mixed-integer linear programming model for the long-term planning of electric power infrastructures considering high renewable penetration. However, all the parameters are deterministic in the model, and their solution approach loses the finite convergence property due to potential duality gap.
The multistage stochastic model described by \cite{LiSiCo2018} considers only continuous decision variables of both capacity expansions and hourly operations. In contrast, our model involves many integer variables, which are important to model assets with significant sunk costs (such as coal and nuclear plants) and to model the flexibility of fossil fuel plants as how fast they can be turned on or shut down. The fine-scale demand and the wind and solar variability in their model are represented via deterministic days in their model, while we consider the stochasticity in each operating period to more accurately capture the effect of short-term operations on long-term investments. 

Our work makes several contributions to the existing literature on power system capacity expansion.
First, we propose a multistage and multiscale stochastic programming model that explicitly takes into account multiple sources of uncertainties, such as investment costs, fuel costs, renewable energy outputs and electricity demand, and captures them in different time scales. The model optimizes the expansion planning of generation, tranmission network and storage simultaneously, and integrates long-term capacity expansion decisions with short-term system operations. The benefit of using our model is demonstrated by computational experiments. 
Second, our model is more computationally challenging compared to many previous studies especially due to the complex stochasticity and many integer variables in the problem. 
We propose an exact algorithm that consists of two layers of decomposition to tackle the problem, and implement it with parallel computation to further improve the performance in terms of reducing total solving time. 
We believe that our modeling and algorithm framework could have broader application areas beyond the energy sector, as long-term decision-making subject to multiscale uncertainties is ubiquitous in the real-world.

\section{Model Description}\label{sec:mdl}
We present in this section the proposed MM-SMIP model that integrates long-term capacity expansion and short-term operation models, followed by the detailed formulation of the constraints that are considered in both the upper and lower levels.
\subsection{The Integration of Long-term Capacity Expansion and Short-term Operations}
\begin{figure}[htbp]
	\begin{center}
		\includegraphics[width=0.7\textwidth]{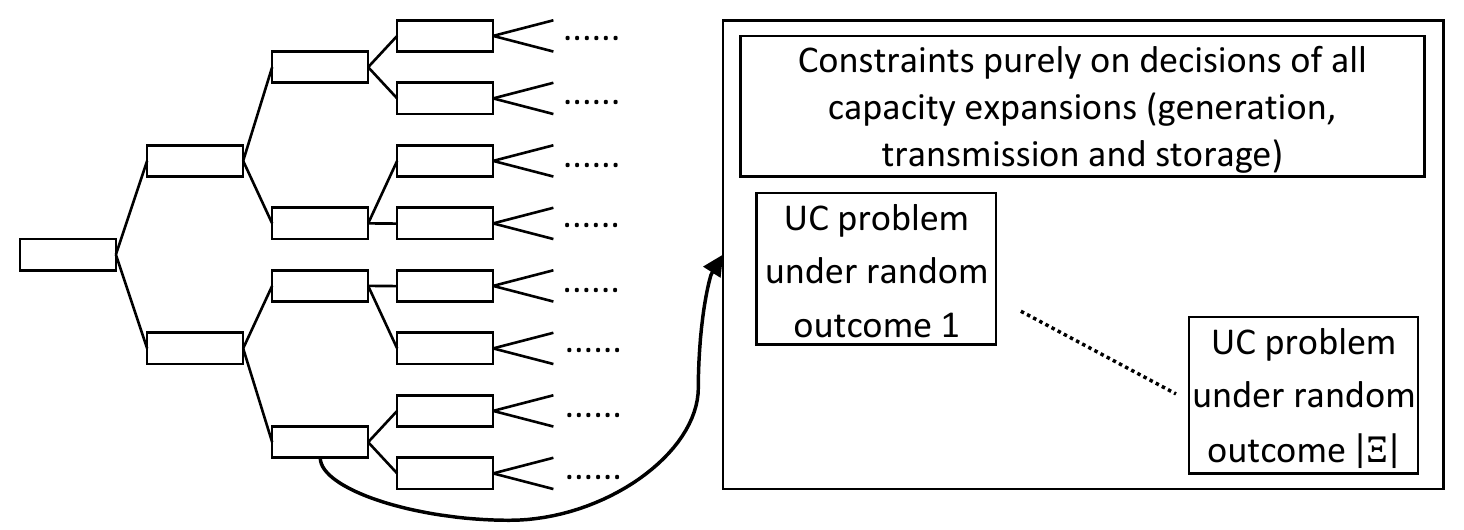}
	\end{center}
	\caption{Hybrid here-and-now and wait-and-see modeling.}\label{fig:rgnP}
\end{figure}

Both the long-term and short-term models are multistage decision problems in nature.
However, the time scales and underlying uncertainties of the long-term and short-term problems are quite different.
Infrastructure expansions are usually planned several years ahead due to their long lead time (development, construction, obtaining permits, etc), and the uncertainties are usually investment costs.
%The planning horizon of an electricity system is even longer, ranging from several years to several decades.
On the other hand, the time scales for unit commitment decisions are in hours (or even in minutes), and the uncertainties are mainly renewable power outputs and real time demand.
A system operator must ensure system reliability by running a unit commitment model repeatedly with updated system information to decide which units to dispatch to meet real-time demand.
%and to manage the variability of electricity generation from intermittent resources.
Since electricity transmission and generation capacity expansions are mainly driven by reliability requirements,
even hourly-scale decisions can have impacts to long-term planning.
Hence, to accurately study the infrastructure needs of an electricity grid to maintain reliability and to support high penetration of variable-output resources,
it is imperative to incorporate detailed unit commitment operations to a long-term model.
The different decision-making time scales lead to the proposed multiscale modeling approach in this paper. The essence is that it combines a here-and-now type of modeling/decision making for the coarse-scale problems (i.e., the investment decisions) and a wait-and-see type of modeling for fine-scale problems (i.e., unit commitment problems), as illustrated in Figure \ref{fig:rgnP}.

In such an approach, long-term expansion decisions are represented in a multistage stochastic mixed integer programming model,
which will provide an expansion profile; namely, an expansion plan corresponding to each of possible future uncertainties in consideration.
Short-term operational costs from unit commitment and economic dispatch are derived from a wait-and-see type of stochastic mixed integer programming model,
which provides the expected costs of various unit commitment schedules along different scenario paths.

\subsection{Coarse-Temporal-Scale Modeling: The Capacity Expansion Model}
In the long-term planning/expansion process, investment decisions at each stage need to be made before future events and uncertainties are realized.
To ensure that our model is useful for decision making (as opposed to what-if type analyses), we enforce the nonanticipativity constraints on the investment decisions; that is, at a given stage, with all the past uncertainties realized, we seek the optimal decisions to minimize the present value of the total system costs, subject to various future uncertainties.
We refer to this problem the upper-level (or strategic-decision level) model in the integrated model to be presented.

\begin{figure}[htb]
	\begin{center}
		\includegraphics[width=0.4\textwidth]{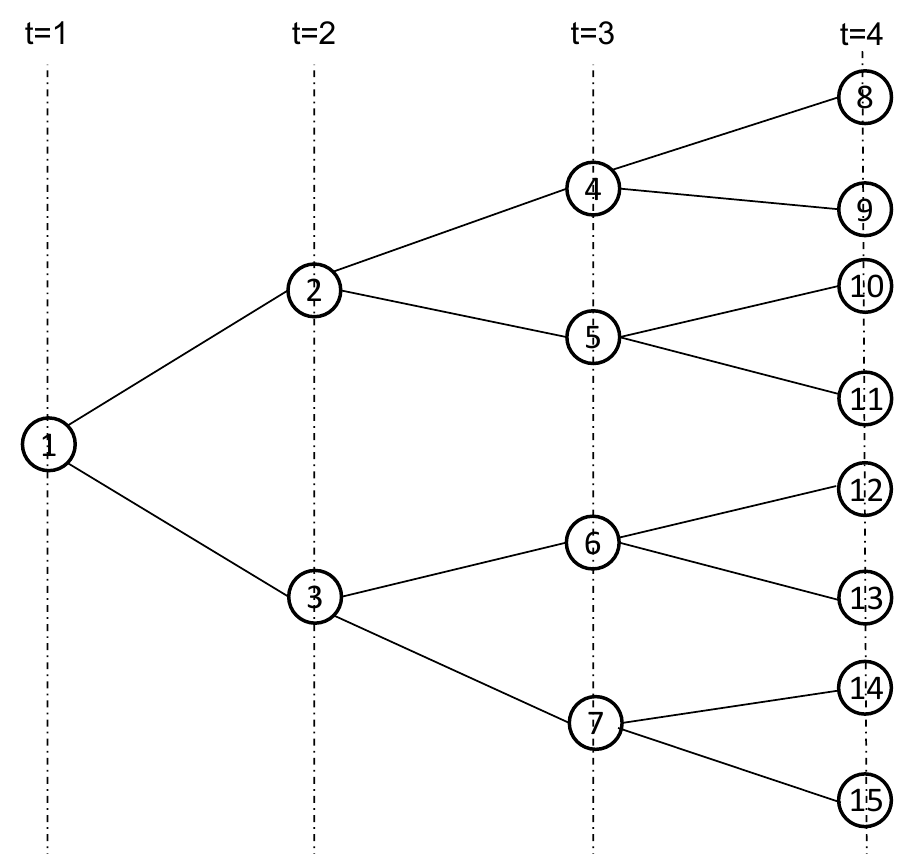}
		\caption{A strategic-level scenario tree describing the here-and-now-type expansion decision making. }
		\label{fig:scenario_tree}
	\end{center}
\end{figure}
The decision variables in the upper level are the expansion (or retrofit) decisions of transmission lines,
electricity power plants (both fossil fuel and renewable plants) and energy storage. Such variables have much more coarse temporal scales when compared with the daily operational decisions in the lower level. The decision-making process in the upper level can be described by a scenario tree, as shown in Figure \ref{fig:scenario_tree}.
In such a tree, we refer to each node as a scenario-tree node, which represents a potential state of the world at the corresponding time $t$, and use $\mathcal{N}$ to denote the set of all such nodes in one scenario tree. In this work, we assume that the set $\mathcal{N}$ is always discrete and finite. While this may be perceived as a restriction as uncertainties in the upper level, such as investment costs, usually have a continuous probability distribution, there have been extensive study on how to construct a scenario tree from continuous distributions and how to assess the quality of the decisions based on a scenario tree. Such issues are outside the scope of this work and they are addressed in, for example, \cite{HeWe09, DeErWe12}.
The capacity expansion decisions are made at each scenario-tree node $n \in \mathcal{N}$ such that the expected value of the total investment cost is minimized over a specific time horizon.

For the ease of argument, we summarize the needed sets notations, parameters, and decision variables in Table
\ref{cep:tab:upper input_var}.
The parameters in the table with a superscript $n$
are random variables that are driven by the underlying (upper-level) uncertainties $\omega \in \Omega_{\omega}$, where $\Omega_{\omega}$ represent the sample space for $\omega$.
The variables all have the superscript $n$, meaning that they are the decision variables corresponding to a scenario-tree node $n \in \Cal{N}$; for example, at $n = 4$ in Figure \ref{fig:scenario_tree}, $x_{g}^{n=4} = 1$ means that power system decides to build the generation unit $g$ when the future world coincides with the state described by the scenario-tree node 4.
\begin{table}[!htbp]
	\caption{Long-Term Related Sets, Indices, Parameters and Decision Variables}
	\begin{center}
	\resizebox{\linewidth}{!}{
		\begin{tabular}{ll}
            \toprule
             \multicolumn{2}{l}{\textbf{Sets and Indices}}  \\
			%$t = 1, \ldots, T$ & time periods (or stages) in the upper-level model\\
			$n\in \Cal{N}$ & a node $n$ in the set of all nodes of a scenario tree, denoted by $\Cal{N}$, and $|\Cal{N}| = N$, \\
			& where $|\cdot|$ denotes the cardinality of a set;  \\
			$j\in \mathbb{J}$ & a bus\footnote{A bus here refers to either the point where a group of power plants connect to the bulk transmission network, or the point where utility companies/consumers withdraw electricity, or both (meaning that there is both generation and consumption at the same bus).} $j$ in the set of all buses in a transmission network, denoted by  $\mathbb{J}$, and $|\mathbb{J}| = J$; \\
		%	a transmission line $l$, connecting bus $i$ and $j$, where $i, j \in \mathbb{J}$, in the set of all arcs in a \\
		%	& transmission network, denoted by $\mathbb{L}$, and $|\mathbb{L}| = L$;\\
			$\hat{j} = 1, \ldots, \hat{R}$ & reserve margin regions; usually a $\hat{j}$ consists a group of buses; \\
			$\hat{J}(j) = \hat{j}$ & $\hat{J}(j)$ is a mapping that maps a bus $j$ to a reserve margin region $\hat{j}$; \\
			$\mathbb{G}_j$ & the set of existing generators at bus $j$; let $\mathbb{G} := \sum_{j \in \mathbb{J}}\mathbb{G}_j$, and $|\mathbb{G}| = G$;\\
			$\mathbb{G}'_j$ & the set of potential generators at bus $j$;
			 let $\mathbb{G}' := \sum_{j \in \mathbb{J}}\mathbb{G}'_j$, and
			 $|\mathbb{G}'| = G'$; \\
            $\mathbb{L}$ & the set of existing transmission lines;\\
			$\mathbb{L}'$ & the set of potential transmission lines; $|\mathbb{L}'| = L'$;\\
            $\mathbb{S}_j$ & the set of existing energy storage devices at bus $j$;let $\mathbb{S} := \sum_{j \in \mathbb{J}}\mathbb{S}_j$, and $|\mathbb{S}| = S$;\\
			$\mathbb{S}'_j$ & the set of potential energy storage devices at bus $j$;
                                let $\mathbb{S}' := \sum_{j \in \mathbb{J}}\mathbb{S}'_j$, and $|\mathbb{S}'| = S'$;\\
			%$u \in \mathbb{U}$ & $u$ can represent a generation unit's type (coal, combined cycle, combustion turbine, \\
%			& wind, solar, etc), or a transmission line type (500 KV AC, 500 KV DC, etc), or a type \\
%			& of energy storage  device (solid state batteries, flow batteries, etc), and $|\mathbb{U}| = U;$ \\
			\hline
            \multicolumn{2}{l}{\textbf{Parameters}}  \\
            $\pi_{n}$ &the probability of reaching scenario-tree node $n$ from the root node; \\
			$c_g^n, c_l^n, c_s^n$ & the cost of building generator $g$, transmission line $l$ and energy storage $s$ \\
            & at scenario-tree node $n$, respectively; [\$]\\
            $a_g, a_l, a_s$ & the capacity of generator $g$, transmission line $l$ and energy storage $s$, respectively; [MW] \\
        	$PK_{\hat{j}}^n$ & peak demand in reserve margin $\hat{j}$ at scenario-tree node $n$; [MW]\\
        	$RM_{\hat{j}}$ & reserve margin requirement for region $\hat{j}$; [\%]\\
        	$DF_{g}$ & derating factors of power plant $g$; [\%]\\
        	%$s = 1, \ldots S$ & possible realizations of uncertainties (i.e., scenarios)\\
        	$p(n)$ & the direct predecessor of scenario-tree node $n$ (assume $a(1) = 0$).\\
			\hline
            \multicolumn{2}{l}{\textbf{Decision Variables}}  \\
            $x^n_g, x^n_l, x^n_s$ & binary variables indicating whether generator $g$, transmission line $l$ and energy storage $s$ are built \\
			& at scenario-tree node $n$, respectively;\\
            $\kappa^{n}_{g}, \kappa^{n}_{l}, \kappa^{n}_{s}$ & available capacity of generator $g$, transmission line $l$, energy storage $s$ \\
            & at scenario-tree node $n$, respectively; [MW]\\
            \bottomrule
		\end{tabular}}
		\label{cep:tab:upper input_var}
	\end{center}
\vspace*{-10pt}
\end{table}
%With a bit abuse of notation, we let $x_n$ denote the collection of all
%$x_{ug}^n$ and $x_{ul}^n$.
%The formulation of the total investment cost (denoted as $TIC$)  at a given scenario-tree node $n\in \Cal{N}$ is as follows:
%\begin{equation}\label{cep:eq:TIC}
%TIC^n (x_n) := \underbrace{\sum_{g\in \mathbb{G}'}\sum_{u\in\mathbb{U}} IC^n_{ug}x_{ug}^n }_{\mathrm{generation/storage\ investment\ costs}}
%+ \underbrace{\sum_{l \in \mathbb{L}'} \sum_{u\in\mathbb{U}} IC^n_{ul}x_{ul}^n}_{\mathrm{transmission\ investment\ costs}}.
%\end{equation}
%
%The time scale between a scenario-tree node and its predecessor is usually in years, and it can be of different lengths over the planning horizon $T$.
%For example, the time interval may be every year at the beginning, with bigger intervals (5-year, 10-year, etc) towards the end of $T$.
The objective of the capacity expansion is to minimize the expected total investment cost over the planning horizon:
\begin{equation*}
	\min \ \sum_{n\in \mathcal{N}}\pi_n \Rm{TIC}^n(x_n)
\end{equation*}
where the cost function with respect to each scenario-tree node $n\in \mathcal N$ is
\begin{subequations}
    \begin{eqnarray}
    \Rm{TIC}^n(x_n) :=
      & \min         & \sum_{g\in \mathbb{G}'} c^n_g x_g^n + \sum_{l\in \mathbb{L}'} c^n_l x_l^n + \sum_{s\in \mathbb{S}'} c^n_s x_s^n  \\
      &\textrm{s.t.} & \kappa_{g}^{n} = \kappa_{g}^{p(n)} + a_g x_g^n,\quad \forall \  g\in \mathbb{G}' \label{cep:eq:DP1-1} \\
	  &              & \kappa_l^n = \kappa_{l}^{p(n)} + a_{l}x_l^n, \quad \forall\ l\in\mathbb{L}' \label{cep:eq:DP1-2} \\
	  &              & \kappa_{s}^{n} = \kappa_{s}^{p(n)} + a_s x_s^n,   \quad \forall\ s\in\mathbb{S}' \label{cep:eq:DP1-3} \\
      &              & \kappa_{g}^{n} \leq a_{g},\ \kappa_{l}^{n} \leq a_{l}, \ \kappa_{s}^{n} \leq a_{s},\quad \forall\ g\in \mathbb{G}',\ l\in\mathbb{L}',\ s\in\mathbb{S}' \label{cep:eq:DP2}\\
      &              & \sum_{j: \hat{J}(j) = \hat{j}}\left(\sum_{g \in \mathbb{G}_{j}} DF_g a_g + \sum_{g\in\mathbb{G}'_j} DF_{g} \kappa_{g}^{n} \right) \geq (1 + RM_{\hat{j}})PK^{n}_{\hat{j}}, \nonumber\\
      &				 & \hspace{2.7in} \forall\ \hat{j} = 1, \ldots, \hat{R}  \label{cep:eq:RA} \\
      &              & \kappa_{g}^{n},\  \kappa_{l}^{n}, \ \kappa_{s}^{n} \geq 0,\ x_g^n,\ x_l^n,\ x_s^n \in \{0, 1\},\quad \forall\ g\in \mathbb{G}',\ l\in \mathbb{L}',\ s\in\mathbb{S}'
    \end{eqnarray}
\end{subequations}

The Constraints \eqref{cep:eq:DP1-1}-\eqref{cep:eq:DP1-3} define that the available capacity at each scenario-tree node $n$ is dependent on the capacity expansion decisions made in its predecessor nodes and those in the current one. As discussed above, the proposed MM-SMIP model imposes a here-and-now multistage structure for the capacity expansion decisions; that is, the expansion decisions need to satisfy non-anticipativity constraints, which is enforced by the tracking parameter $p(n)$. Assume that at the root node $n=1$, $p(1)=0$, and $\kappa_g^0 = \kappa_l^0 = \kappa_s^0 = 0$, for all $g\in\mathbb{G}'$, $l\in\mathbb{L}'$ and $s\in\mathbb{S}'$.
%In constraint \eqref{cep:eq:DP1}, the $\kappa$ variables capture the cumulative capacity built over time; while the incremental capacity addition is represented by the terms $a_{ug}x_{ug}^{n}$ and $a_{ul}x_{ul}$, which reflect the lumpiness of capacity expansions.
%For example, a new coal plant may be built in the increment of 100MW, such as a new plant of 200 MW, 300 MW, etc. It is either technologically impossible or uneconomic to build, say, a 3 MW coal plant.
%where the dynamic programming equations \eqref{cep:eq:DP1} describe the capacity expansion accounting process over the time horizon, and t
Constraint \eqref{cep:eq:DP2} restricts that each infrastructure can be constructed at most once through the planning horizon. Constraint \eqref{cep:eq:RA} reflects a typical resource adequacy requirement. It says that the total installed capacity (multiplying a derating factor) in each reserve region must exceed a certain percentage of the projected peak demand in the region. The certain percentage ($RM_{\hat{j}}$) is called the reserve margin requirement, and a typical value is 15\%. The derating factor is to reflect the fact that the nameplate capacity of a power plant may not be fully available in real-time (such as wind plants).

\subsection{Fine-temporal-Scale Modeling: The Unit Commitment Model}
\label{sec:UC_ED}
%Note that we only present the most basic upper level constraints here, since the main focus of this work is algorithm development, as opposed to providing detailed market analysis using real system data. More constraints can certainly be added to the upper level of the model should they be needed. The following section provides the detailed UC model formulation at each scenario-tree node.
%\subsubsection{Here-and-now vs. wait-and-see operational models}
Within each scenario-tree node, we consider the unit commitment (UC) and economic dispatch (ED) process with much finer decision epochs (usually by hours).
The UC process is to determine the most economic schedule to commit units (that is, to schedule units to be turned on or off), subject to certain reliability requirements, to meet projected demand in real time. The ED process matches electricity supply and demand in real time, based on the pre-determined units' on/off schedules from the UC process. We refer to the UC/ED model as the lower-level (or operation-level) model (and only use ``UC" to refer to such a model in the rest of the paper for simplicity, unless we want to emphasize the different decisions in UC and ED).

Through the capacity expansion planning process, it is neither practical nor necessary to optimize the UC operation over all the hours at a scenario-tree node (for example, 8760 hours in a year).
In fact, system operators in real-world typically solves the UC problem on a rolling basis from one-hour to at most one-week ahead (e.g., \cite{CAISO_RUC}).
To reduce computational burden, power system operations are commonly
accounted for in the capacity expansion using representative time periods of the planning horizon such as hours, days, or weeks \citep{MunozMills2015,Dvorkin2017,Pineda2018}. This time-period aggregation approach is sometimes arguable as it is not able to fully capture long- or mid-term dynamics of renewable power generation and electricity demand. However, such drawback can be overcomed by using sophisticated clustering approaches maintaining the chronology of varying parameters throughout the whole planning horizon \citep{TEICHGRAEBER20191283}.
Our model considers daily UC problems for the operational level with respect to a set of representative days as the operating periods between investments. The representative days are selected so as to capture different effects of lower-level uncertainties. For example, the weather conditions would vary significantly between seasons, resulting in different probability distributions of renewable energy outputs; and the load profiles may also differ drastically between weekdays and weekends.
Let $K_n$ denote the set of representative days in the time period between scenario-tree node $n$ and its immediate successors. For example, in Figure \ref{fig:scenario_tree}, the representations of operating periods between $t=1$ to $t=2$ are included in $K_1$.
%Then at each scenario-tree node $n$, we have $|K_n|$ UC optimization problems to solve.

The system operation in each representative day faces an array of uncertainties including, but not limited to, demand fluctuation, forecasting errors of wind and solar plants' outputs, and forced outage of power plants and/or transmission lines. 
It is natural to model the UC/ED process as a two-stage stochastic program with recourse, with the UC decisions in the first-stage and the ED decisions as
recourse decisions to balance real-time supply and demand.
However, it may not be completely necessary to know
exactly which unit is operating at a particular hour many years from today. Instead, we are interested to know what an economic and robust capacity mix are like in the future that can withstand various real-time uncertainties. 
As a result, we propose to model the process as a wait-and-see-type stochastic program.
%We aim to enhance the long-term planning decisions by capturing, through embedded a wait-and-see-type stochastic UC program within each upper-level scenario-tree node, their impact on real-time operating procedures in the power system.
The wait-and-see feature means that both the UC and ED decisions are made after all the uncertainties are resolved; that is, a deterministic UC problem is solved with respect to each possible realization of uncertainties in a day. 

\begin{table}
	\caption{Parameters, Sets, Indices and Decision Variables for the Short-Term UC Model}
	\begin{center}
	\resizebox{\linewidth}{!}{
		\begin{tabular}{ll}
            \toprule
            \multicolumn{2}{l}{\textbf{Sets and Indices}}  \\
			$H_k$ & the set of hours in day $k \in K_n$, for a scenario-tree node $n\in\mathcal{N}$;  \\
            $(i, j)\in \mathbb{A}$ & the set of O-D pairs $(i, j)$ with an existing transmission line network;\\
			$(i, j)\in \mathbb{A}'$ & the set of O-D pairs $(i, j)$ such that new transmission lines can be built; \\ \hline
            \multicolumn{2}{l}{\textbf{Parameters}}  \\
			$SC_g$ & start-up cost of generation unit $g$; [\$]\\
			$GC_g^h(\cdot)$ & the variable cost function; [\$] \\
			$P_g^{\text{min}}$ & minimum power for generator $g$ to be economically running ($g\in \mathbb{G}\cup \mathbb{G}' $); [MW]\\
            $P_g^{\text{max}}$ & maximum power that can be generated from generator $g$ ($g\in \mathbb{G}\cup \mathbb{G}' $); [MW]\\
			${SR}_{j}^h$      & spinning reserve requirement of bus $j\in \mathbb{J}$ at a particular hour $h$\\
			$RU_{g}$        & ramp-up rate of power generator unit $g$ ($g\in \mathbb{G}\cup \mathbb{G}' $); [MW/hour] \\
			$RD_{g}$        & ramp-down rate of power generator unit $g$ ($g\in \mathbb{G}\cup \mathbb{G}' $); [MW/hour] \\
		    $L_g$ &         minimum run time of generator $g\in \mathbb{G}\cup \mathbb{G}'$; [hour]\\
			$D_{j}^{h}$ & real-time energy demand at bus $j\in \mathbb{J}$ in hour $h$; [MWh]\\
			$F_l$ & transmission capacity of existing transmission line $l\in\mathbb{L}$; [MW]\\
			%$\Delta_{ij}$        & susceptance of the line $(i,j)$, which equals the reciprocal of the reactance of line $(i,j)$\\
			B$_{ij}$ & percentage of energy loss of transmitting energy from bus $i$ to $j$;\\
			$E_s$ & withdrawal efficiency for storage unit $s \in \mathbb{S}\cup \mathbb{S}'$; [\%] \\
			\hline
            \multicolumn{2}{l}{\textbf{Decision Variables}}  \\
            $\alpha_{g}^h$       & the commitment status of generation unit $g$ in hour $h$; $\alpha_{g}^h \in \{0,1\}$\\
			$\gamma_{g}^h$    & whether to turn unit $g$ on or not at the starting of hour $h$; $\gamma_{g}^h \in \{0,1\}$\\
			$p_{g}^h$     & energy generated from unit $g$ in hour $h$; [MWh]  \\
			$s_{g}^h$    & spinning reserve of generator unit $g$ in hour $h$; [MWh]\\
			%$q_j^h$  & actually power utilized from bus $j$ at hour $h$\\
			$u_s^h$  & energy withdrawal from storage device $s $ in hour $h$; [MWh]\\
			$v_s^h$  & energy injection to storage device $s $ in hour $h$; [MWh]\\
			$r_s^h$  & remaining energy in storage device $s $ in hour $h$; [MWh]\\
			$f_{ij}^h$         & energy flow between bus $i$ and bus $j$ in hour $h$; [MWh]\\
		 \bottomrule
		\end{tabular}}
		\label{cep:tab:lower input_var}
	\end{center}
\end{table}

To emphasize the difference of uncertainties in the upper versus in the lower level, we use another letter $\mu$ to represent all lower level uncertainties, and denote its sample space as $\Omega^{n_k}_{\mu}$, where $n\in\Cal{N}$ and $k\in K_n$. The superscript $n_k$ means that at different scenario-tree nodes or in different operating periods, the sample spaces of the lower-level uncertainty may be different. For example, consider again the scenario tree in Figure \ref{fig:scenario_tree},
at $t =2$, the node $n=2$ may represent a state that the economy is booming (with lower investment costs due to easier access to capital); while $n=3$ may represent a state of a sluggish economy (and higher capital costs). Then the distribution of real-time demand is likely different at $n=2$ than at $n=3$, with the demand at $n=2$ expected to be higher on average. Further let $\zeta_{n_k}(\mu)$ be a random vector that is the collection of all random variables at the lower level in day $k\in K_n$, which are all defined on the probability space  $(\Omega^{n_k}_{\mu}, \Sigma^{n_k}_{\mu}, P^{n_k}_{\mu})$. Also let $\Xi^{n_k}$ denote the support of $\zeta_{n_k}$.
We present the UC problem with respect to the upper-level decision $x_n$ and the lower-level random vector $\zeta_{n_k}$ in the following, where the parameters and decision variables related to the UC process are summarized in Table \ref{cep:tab:lower input_var}.
\begin{subequations}\label{DUC}
	\begin{eqnarray}
	& \min          &  \Rm{TOC}_n^k(y_n^k(\zeta_{n_k})) = \sum_{h\in H_k}\sum_{g \in \mathbb{G}\cup \mathbb{G}'}\left[SC_{g}\gamma_{g}^{h} + GC_g^h(p_{g}^{h}, \ s_g^h)\right] \label{uc:obj}\\
	& \mathrm{s.t.} &  \alpha_{g}^h\leq {\frac{\kappa_{g}^{n}}{a_{g}}}, \quad \forall g\in \mathbb{G}',\ h\in H_k, \label{uc:avai}\\
	&               &\alpha_{g}^\tau-1\leq \alpha_{g}^h-\alpha_{g}^{h-1}\leq \alpha_{g}^\tau, \ \tau = h,\dots, \min \{h+L_g-1, |H_k|\}, \nonumber \\
	&               & \hspace{1.9in} g \in \mathbb{G} \cup \mathbb{G}', h\in H_k, \label{uc:minup}\\
	&               &\gamma_{g}^h\geq \alpha_{g}^h-\alpha_{g}^{h-1}, \quad \forall g \in \mathbb{G} \cup \mathbb{G}',\  h = 2, \ldots, |H_k|,\label{uc:strtup}\\
	&               &p_{g}^h+s_{g}^{h}\leq
                    	\begin{cases}
                        	P_g^{\max}\alpha_{g}^h, & \forall  g \in \mathbb{G},\ h \in H_k,\\
                        	\kappa_{g}^{n}, & \forall g\in \mathbb{G}',\ h\in H_k,
                    	\end{cases}\label{uc:spin}\\
	&               &P^{\text{min}}_g\alpha_{g}^h\leq p_{g}^h,\quad  \forall g\in \mathbb{G}\cap \mathbb{G}',\ h \in H_k,\label{uc:genLim}\\
	&               &\sum_{g \in G_j\cup G'_j} s_{g}^{h} \geq SR_{j}^{h}, \quad \forall j \in \mathbb{J},\ h \in H_k, \label{uc:spinlim}\\
	&               &p_{g}^{h} - p_{g}^{h-1}\leq RU_{g}, \quad p_{g}^{h-1} - p_{g}^{h}\leq RD_{g}, \quad \forall  g \in \mathbb{G} \cap \mathbb{G}',\ h = 2, \ldots,  |H_k|,\label{uc:rampup}\\
	&               & r^{1}_s = 0,\ r^h_s = r^{h-1}_s + E_sv_s^{h-1}-u_s^{h-1}, \quad  s\in \mathbb{S}\cup\mathbb{S}',\ h=2, \ldots, |H_k|, \label{uc:strpwr}\\
	&               &u^h_s \leq r^h_s, \quad \forall s\in\mathbb{S}\cup\mathbb{S}',\ h\in H_k,\label{uc:strpwrrls}\\
	&               & r^h_s\leq \kappa_{s}^{n}, \quad  s\in \mathbb{S}',\ h\in H_k, \label{uc:storagelim}\\
	&               & D_j^h + \sum_{i: (j, i) \in A} f_{ji}^h - \sum_{i: (i, j)\in A}(1 - B_{ij})f_{ij}^h \  =
                            \sum_{g \in  \mathbb{G}_j\cup \mathbb{G}'_j}p_{g}^{h} + \sum_{s \in  \mathbb{S}_j\cup \mathbb{S}'_j} \left(u^h_s - v_s^h\right), \nonumber \\
    &               &      \hspace{18em}\forall  j \in \mathbb{J}, h\in H_k, \label{uc:dcKCL} \\
	%&               &\left| f_{ij}^h-\Delta_{ij}\left(\theta_{i}^h - \theta_{j}^h\right) \right|\leq
%                    	\begin{cases}
%                    	0, & \forall\left(i,j\right)\in \mathbb{A}, h\in H_k,\\
%                    	M(1-\phi_{ij}^{n}), & \forall (i,j)\in \mathbb{A}', h\in H_k\\
%                    	\end{cases}\label{uc:dcKVL}\\
	&               &f_{ij}^h \leq \kappa_{ij}^{n}, \quad  \forall\left(i,j\right)\in \mathbb{A}\cup \mathbb{A'}, h\in H_k  \label{uc:flowlim}\\
	&               &\alpha_{g}^h,\ \gamma_{g}^h\in \{0,1\},\quad \forall g \in G_j\cup G'_j, j \in N,  \forall j \in N,  h \in H_k,\label{uc:Bin}\\
	&               &p_g^h,\ s_g^h,\ r_s^h,\ u_s^h,\ v_s^h,\ f_{ij}^h\geq 0, \quad \forall g\in \mathbb{G}\cup\mathbb{G}',\ (i, j)\in \mathbb{A}\cup\mathbb{A}',\ s\in \mathbb{S}\cup\mathbb{S}',\ h\in H_k
	\end{eqnarray}
\end{subequations}

In the above problem, the temporal resolutions for decision making are hours, hence the $h$ superscripts to all the decision variables. The UC-related decision variables are $\alpha^h_g$ 
%-- denoting the status of unit $g$ in hour $h$ (with 0 meaning $g$ is off in $h$, and 1 meaning it is on), 
and $\gamma^h_g$.
%-- indicating to turn on unit $g$ or not at the beginning of hour $h$.
The other variables (as listed in Table \ref{cep:tab:lower input_var}) are associated with the ED process.
In objective function \eqref{uc:obj}, the first part corresponds to the generic UC cost function which gives the total costs in the representative day $k$ associated with start-up costs;\footnote{We assume it is free to turn off a unit here. Turn-off costs can be easily added to the UC model if needed.} the second part of \eqref{uc:obj} corresponds to the VOM costs. The individual VOM cost function $GC_g^h(\cdot)$ is usually a quadratic function of positive second derivative (hence, convex). If it is preferred to solve a mixed integer linear program (MILP) as opposed to a mixed integer quadratic program (MIQP) when solving each of the deterministic UC problem, we can use a piecewise linear function to replace the quadratic cost function (cf. \cite{HoRoONCh01,ZhWaPaGu11}).

The first constraint \eqref{uc:avai} in the UC model is to ensure that if a potential generation unit $g$ has not been built until scenario-tree node $n$, it cannot be available for dispatch for any of the hours under the node $n$; that is, $\alpha_g^h = 0$ for all $h\in H_k$ and $k\in K_n$.
The minimum up and down time constraints of each generator are defined in \eqref{uc:minup}. The constraint \eqref{uc:strtup} is to determine whether unit $g$ is to start or turn off in hour $h$.
The upper bound on energy generated and spinning reserve service provided by uint $g$ is defined by constraint \eqref{uc:spin}, and the lower bound of energy generation by unit $g$ (if it is running) is defined by constraint \eqref{uc:genLim}. Constraint \eqref{uc:spinlim} ensures that the spinning reserve for each bus exceeds a required amount. The change in power outputs of each generator between hours is restricted by its maximum ramping up and ramping down rates, as defined in \eqref{uc:rampup}.
In \eqref{uc:strpwr}, The coefficient $E_s$ in front of the charging variable $v_s^h$ is to capture efficiency loss of any storage technology; namely, for each one MWh energy charged into a storage device, only a fraction of the one MWh (e.g., $E_s = 80\%$) are available for withdrawal later).
Equation \eqref{uc:dcKCL} is the DC approximation of Kirchhoff's Current Law (KCL), which is exactly a mass balance constraint for energy injecting (total energy produced at the bus plus energy transmission into the bus) and withdrawing (demand, storage charging, and energy transmission out of the bus) from each bus $j\in\mathbb{J}$. Constraint \eqref{uc:flowlim} defines that the power flow through a transmission line is limited by its capacity.
The upper-level capacity expansion decisions and lower-level operational decisions are explicitly linked through the constraints \eqref{uc:avai}, \eqref{uc:spin}, \eqref{uc:storagelim} and \eqref{uc:flowlim}, with the real-time energy generation, storage and transmission limited by the total installed capacity at a particular scenario-tree node $n$. 
%In addition, the upper level decisions are implicitly linked to the minimum-up/down time constraints \eqref{uc:minup} and ramping constraints \eqref{uc:rampup}, as the types of generation units added to the system in the upper level, along with the existing capacity, will determine how flexible the system is in response to real-time uncertainties.

The uncertainties at the operational level can include fuel costs in the objective function \eqref{uc:obj} and real-time demand ($D_j^h$) in \eqref{uc:dcKCL}. While the UC formulation above does not explicitly model variable-output resources, such as wind and solar, their output variability can be easily incorporated by timing a random variable to their capacities in \eqref{uc:spin}.
With lower level uncertainties (especially in demand and wind/solar plants' outputs), it is possible that under certain scenarios, the UC problem is infeasible. To prevent this from happening, we can assign a large VOM cost to one generator at each bus, and do not assign a capacity bound on the generators; in another words, we let the most expansive plant at each bus represent the (nodal) unserved energy, and the large VOM costs correspond to consumers' value of lost load. In this way, the overall stochastic program has a \emph{relatively complete recourse}; that is, for any upper level decision $x\in \mathbb{X}$, the lower level parameterized set $\mathbb{Y}(x, \zeta_{n_k})$ is feasible for any $\zeta_{n_k}\in\Xi^{n_k}$, for all $n\in \mathcal{N}$, and $k\in K_n$.

\subsection{A Complete, Compact Formulation}
%{multistage and multiscale stochastic mixed integer programming model}
%As described above, our proposed problem combines a hear-and-now capacity expansion model at the upper level and a set of weekly wait-and-see UC models at the lower level. The expansion planning (strategic level) is a multistage stochastic mixed integer program in its own right. The random process of its here-and-now decision making can be described by a scenario tree, as shown in Figure \ref{fig:scenario_tree}.
%In such a tree, we refer to each node as a scenario-tree node, which represents a potential state of the world at the corresponding time $t$, and use $\mathcal{N}$ to denote the set of all such nodes in the scenario tree. The infrastructure expansion decisions are made within each node $n \in \mathcal{N}$ such that the expected total cost (including investment and operational cost) is minimized over a specific time horizon. The feasibility and corresponding operational cost of the capacity expansion decisions are examined by the stochastic UC model embedded within each scenario-tree node. In each node, the stochastic UC problem include a series of parallel deterministic UC models as in \eqref{DUC}, each of which is not only linked with the capacity variables but also unique by itself in terms of fuel costs, renewable energy supplies and electricity demands. This series of scenarios of the lower-level uncertainty are simulated by the Monte Carlo method.

To facilitate the discussion of algorithm development in the next section, we present a complete, but compact formulation of the overall multistage and multiscale stochastic mixed integer programming (MM-SMIP) model as follows,
\begin{subequations} \label{eq:smip}
	\begin{eqnarray} \textrm{[MM-SMIP]:}
	&\min\limits_{x,\ y}
	& \sum_{n\in \mathcal{N}}\pi_n \left[\Rm{TIC}^n(x_n)+\sum_{k\in K_n}\mathbf{E_{\zeta_{n_k}}} \min_{y_n^k}\ \Rm{TOC}(y_n^k(\zeta_{n_k}))\right] \label{eq:smip:split}\\[4pt]
	&\textrm{s.t.}&  B y_n^k(\zeta_{n_k}) \leq V(\zeta_{n_k})(b + \sum_{m\in \mathcal{P}_n}A_m x_{m}), \nonumber \\
	&			  &  \hspace{60pt} \forall\ n\in \mathcal{N},\ k\in K_n,\  \zeta_{n_k} \in \Xi^{n_k} \label{eq:smip:coupling} \\
	&             &  x_n \in X_n \cap \{0,1\}^{G'+L'+S'}, \quad \forall n\in \mathcal{N} \label{eq:smip:upper_constr}\\
	& 	       &y_n^k(\zeta_{n_k}) \in Y_n^k(\zeta_{n_k}), \quad \forall n\in \mathcal N, k\in K_n,\ \zeta_{n_k} \in \Xi^{n_k}. \label{eq:smip:operation}
	\end{eqnarray}
\end{subequations}
%where $\pi_n$ is the absolute probability of $n$ (not the transition probability from $n$'s immediate predecessor to $n$). $c_n$ is the vector of investment costs in the objective function.
where \eqref{eq:smip:coupling} represents the linkage constraints \eqref{uc:avai}, \eqref{uc:spin}, \eqref{uc:storagelim} and \eqref{uc:flowlim}; $B$, $V(\zeta_{n_k})$ and $b$ are appropriate matrices and vectors coefficients derived from the constraints.
The constraint \eqref{eq:smip:upper_constr} represents the upper-level constraints, and the lower-level constraints (minus all the linkage constraints) are included by the set $Y_n^k(\zeta_{n_k})$ in \eqref{eq:smip:operation}. Note that without the linkage constraints, the rest of the lower-level constraints do not contain any upper-level variable; and hence the set $Y_n^k(\zeta_{n_k})$ does not depend on $x$. 

Using a hybrid here-and-now (for the upper level) and wait-and-see (for the lower level) modeling is a salient feature of our proposed multiscale stochastic optimization.
This approach actually resembles to industry practices where utility companies or system operators first identify infrastructure expansion plans, and then conduct feasibility analysis to test if such plans can meet real-time demand under various scenarios. Our approach combines such separate (and often iterative) processes into an integrated model, and we believe such an approach can enjoy applications in many other areas as well.

% The matrix $B$, $V(\zeta_{n_k})$ and the vector $b$ in \eqref{eq:smip:coupling}, all of proper dimensions, are given parameters.
% The matrix $B$ is just a matrix of 0's and 1's to indicate which specific lower-level variables are included in which specific constraints; $V(\zeta_{n_k})$ represents the (random) availability percentage of the corresponding capacity in real-time;
% while the vector $b$ may represent existing capacity.
%unlike the generic lower-level constraint set $\mathbb{Y}(x, \zeta_{n_k})$ in \eqref{DUC}.

\section{Solution Algorithms}\label{cep:sec:alg}
The MM-SMIP model presented in the previous section will always lead to large-scale mixed integer problems when applied to real-world-scale systems.
The total number of integer variables in the upper level of the model is $N (G'+ L' + S')$, and in the lower level is $2\sum_{n\in \mathcal N}\sum_{k\in K_n}(S^k_n|H_k|)(G + G')$, where $S^k_n$ is the number of scenario paths at scenario-tree node $n$, within day $k$, that are generated via Monte Carlo simulation.
%To provide a concrete example, to plan for infrastructure expansion on an IEEE 57-bus system (with 7 existing generators) within the next 5 years (of yearly resolution at the upper level), the resulting model will have more than 50 million integer variables if we generate 100 lower-level scenarios in each week at each scenario-tree node.
With a much large system, it is unlikely that the state of the art commercial mixed integer optimization solvers, such as CPLEX and Gurobi, can be applied to solve MM-SMIP directly. (This is indeed the case as demonstrated in our numerical experiments within an IEEE 118-bus test system. The details are provided in Section \ref{sec:exp}.)
We discuss in this section our proposed NCD algorithm that consists of two layers of decomposition and fully exploits the structure of the MM-SMIP to facilitate massively parallel computing. 
While the decomposition method we use to deal with the upper-level expansion decisions is directly inspired by that in \cite{SiPhWo09}, it is enhanced significantly in order to handle the multiscale strucutre of our model. 
It will be shown in Section \ref{sec:exp} that the standard column generation approach as applied by \cite{SiPhWo09} fails to address the computational difficulty even in a small power system. 

\subsection{Dantzig-Wolfe Decomposition for Solving Here-and-Now Expansion Planning}\label{subsec:cg}
A key aspect to note for the upper-level problem is that if the stage-linking constraints \eqref{eq:smip:coupling} are excluded, the remaining problem is separable by scenario-tree nodes, which is referred to as nodal decomposition.
We can apply a reformulation technique known as the variable splitting (cf. \cite{AhKiPa03,SeYuGe06,SiPhWo09}) to eliminate the needs of having the cumulative capacity variables $\kappa_g$ and $\kappa_l$, and hence, to enable nodal decomposition. An additional benefit of such reformulation is that for the resulting mixed-integer problem, by relaxing the integrability constraints (of the upper level problem),
its optimal objective value provides a tighter lower bound than directly relaxing integrability constraints of the original problem, as shown in (\cite{SiPhWo09}).
Based on the concept of scenario-tree paths, we can introduce auxiliary variables $z_{n}$ to represent whether the expansion of an infrastructure's capacity has been made along the scenario path $\mathcal{P}_n$. With such notation, we present the following equivalent reformulation of the [MM-SMIP] model:
\begin{subequations} \label{eq:mp}
	\begin{eqnarray}
	&\min\limits_{x, \ z}
	& \sum_{n\in \mathcal{N}}\pi_n \left[\Rm{TIC}^n(x_n)+\sum_{k\in K_n}\mathbf{E}_{\zeta_{n_k}} \min_{y_n^k}\ \Rm{TOC}(y_n^k(\zeta_{n_k}))\right] \\
	&\textrm{s.t.}& z_{n} \leq \sum_{m\in \mathcal P_n}x_m, \quad \forall n\in \mathcal{N} \label{eq:mp:split1} \\
	&			  & \sum_{m\in \mathcal P_n}x_m \leq \v{1}, \quad \forall n\in \mathcal{N} \label{eq:mp:split2} \\
    &             &   B_n^ky_n^k(\zeta_{n_k}) \leq V(\zeta_{n_k})(b_n^k + A_nz_n), \quad \forall n\in \mathcal{N},\  k \in K_n,\ \zeta_{n_k} \in \Xi^{n_k}  \label{eq:mp:coupling} \\
	&             &  x_{n} \in X_n\cap \{0,1\}^{G'+L'+S'},  \quad \forall n\in \mathcal{N} \\
	&             & z_n \in \{0,1\}^{G'+L'+S'}, \quad \forall n\in \mathcal{N} \\
	& 	       &y_n^k(\zeta^k) \in Y_n^k(\zeta^k), \quad \forall n\in \mathcal N, k\in K_n, \zeta^k\in \Xi^k. \label{eq:mp:operation}
	\end{eqnarray}
\end{subequations}

%\subsubsection{Column generation}
%\label{subsec:CG}
With the introduction of the auxiliary variables $z_{n}$, our problem can be decomposed into a master problem and a set of subproblems, aka the Dantzig-Wolfe decomposition, and we apply the column generation approach to solve the decomposed problems.
A key aspect for the decomposition scheme is the fact that since expansion decision variables are binary, the number of possible outcomes for the system capacity within each scenario-tree node is finite (though it can be extremely large). We use the index $i$ to label such possible capacity outcomes, and let the set $\mathcal{F}_n$ denote the collection of such indices. Then the upper-level primary master problem, denoted as [PMP], can be written as follows:
\begin{subequations}
	\begin{eqnarray} \textrm{[PMP]:} \quad
	&\min\limits_{x,\ \lambda} \quad
	& \sum_{n\in \mathcal{N}}\pi_n\left[ \Rm{TIC}^n(x_n) +  \sum_{i\in \mathcal {F}_n}\sum_{k\in K_n}(\overline{\Rm{TOC}}_n^k)^i \cdot \lambda_n^i \right] \\
	&\textrm{s.t.} \quad
	&  \sum_{i\in \mathcal {F}_n}z_{n}^i\lambda_n^i \leq \sum_{m\in \mathcal P_n} x_m,   \quad \forall n\in  \mathcal{N}, \quad \leftarrow \psi_{n} \\
	&			  & \sum_{m\in \mathcal P_n}x_m \leq \v{1}, \quad \forall n\in \mathcal{N} \\
	&             & \sum_{i\in \mathcal {F}_n}\lambda_n^i = 1 \quad \forall n\in  \mathcal{N}, \quad \leftarrow \psi_n^0  \label{eq:pmp:conv}\\
	&             & x_n\in X_n\cap \{0,1\}^{G'+L'+S'}, \ \lambda_n^i \in \{0,1\} \quad \forall n\in \mathcal{N},\ i\in \mathcal F_n.
	\end{eqnarray}
\end{subequations}
In [PMP], the binary variables, $\lambda_n^i$, indicate whether the capacity outcome $i \in \mathcal{F}_n$ is applied (1: yes; 0: no); and the convexity constraints \eqref{eq:pmp:conv} ensure that the capacity expansions along the scenario path $\mathcal P_n$ yields exactly one capacity outcome for each scenario-tree node. Once the system capacity $i \in \mathcal{F}_n$ is determined, the lower level UC problem can be solved, and the notation $(\overline{\Rm{TOC}}_n^k)^i$ represents the corresponding expected value of the optimal VOM cost in day $k\in K_n$.
%The structure of [PMP] is illustrated in Figure \ref{fig:rpmp}, with each block representing a possible expansion plan at the corresponding scenario-tree node.
%\begin{figure}[!htb]
%	\centering
%	\includegraphics[scale=0.5]{figures/rpmp}
%	\caption{The Block Structure of the Primary (Strategic) Master Problem [PMP]}
%	\label{fig:rpmp}
%\end{figure}
Let [MM-SMIP]$^\ast$ and [PMP]$^\ast$ denote the optimal objective function value of the original [MM-SMIP] (aka, Problem \eqref{eq:smip}) and the primary master problem [PMP], respectively. It is not difficult to see that [MM-SMIP]$^\ast = $ [PMP]$^\ast$.

To solve [PMP], we first perform a continuous relaxation of the integer requirements for the variables $x_n$ and $\lambda_n^i$, obtaining a linear primary master problem [PMP-LR]. We apply the classic column generation approach to solve [PMP-LR] and use parallel computation wherever possible to handle the extremely large problem size.
%The optimal objective function value, [PMP-LR]$^\ast$, would expect to provide a lower bound for [PMP]$^\ast $.
Note that the number of possible system capacity expansion results for each scenario-tree node $n$, i.e.,  $|\mathcal {F}_n|$, would be extremely large,
especially when there are many sites that can build new capacity and many technology types to choose from. For example, $|\mathcal {F}_n| = 2^{20}$ if there are $20$ potential generators to build, and the number would exponentially increase further when the expansion of transmission lines and storage facilities are taken into account.
% as one can choose to apply various expansion approaches and increase the capacity for different infrastructures in any decision stage.
It is neither applicable nor necessary to include all the columns $\lambda_n^i,~\forall n\in \mathcal{N}, i\in \mathcal {F}_n$ in  s [PMP-LR]. With the column generation algorithm, one can instead consider only a subset of these columns, yielding a restricted formulation denoted as [RPMP-LR], and pricing out more columns iteratively by solving a set of subproblems with respect to the scenario-tree nodes.
%(denoted as $\widetilde{\mathcal{F}}_n$, with $\widetilde{\mathcal{F}}_n \subset \mathcal{F}_n$ )
%The formulation of [RPMP-LR] is as follows:
%\begin{subequations}\label{RPMP_LR}
%	\begin{eqnarray} \textrm{[RPMP-LR]:} \quad
%	&\min\limits_{x,\ \lambda} \quad
%	& \sum_{n\in \mathcal{N}}\pi_n\delta^{t(n)}\left[ \Rm{TIC}^n(x_n) +  \sum_{i\in \widetilde{\mathcal{F}}_n}\sum_{k\in K_n}(\overline{\Rm{TOC}}_n^k)^i \cdot \lambda_n^i \right] \\
%	&\textrm{s.t.} \quad
%	& \sum_{i\in \widetilde{\mathcal{F}_n}}z_{n}^i\lambda_n^i \leq \sum_{m\in \mathcal P_n} x_m,   \quad \forall n\in  \mathcal{N}, \quad \leftarrow \psi_{n}  \\
%	&			  & \sum_{m\in \mathcal P_n}x_m \leq \v{1}, \quad \forall n\in \mathcal{N} \\
%	&             & \sum_{i\in \widetilde{\mathcal{F}}_n}\lambda_n^i = 1 \quad \forall n\in  \mathcal{N}, \quad \leftarrow \psi_n^0 \\
%	&             & x_n \in X_n \cap \mathbb{R}^{G'+L'+S'}, \quad \forall n\in \mathcal{N}\\
%	&			&   \lambda_n^i \geq 0,\  \forall n\in \mathcal{N}, i\in \widetilde{\mathcal{F}}_n.
%	\end{eqnarray}
%\end{subequations}
%In [RPMP-LR], we use $\psi_{n}$ and $\psi_n^0$ to denote the Lagrangian multipliers of the corresponding constraints.
%The subset $\widetilde{\mathcal{F}}_n$ can be initialized with the most ``expensive" columns that have every facility expanded as much as possible at each period along the time horizon until the maximum capacity limit is reached. %We assume that this initial problem is feasible, as otherwise the original problem would be infeasible.
Given a dual optimal solution $(\hat\psi_{n}, \hat \psi_n^0)$ of the [RPMP-LR], the pricing subproblem with respect to the scenario-tree node $n$, denoted as $\textrm{[PSP}_n\textrm{]}$, is shown as follows, 
\begin{subequations}
	\begin{eqnarray} \textrm{[PSP}_n\textrm{]}: \quad
	&\min\limits_{z,\ y} \quad&  \pi_n \sum_{k\in K_n}\mathbf{E_{\zeta_{n_k}}} \left[\min_{y_n^k}\  \Rm{TOC}(y_n^k(\zeta_{n_k})) \right] - \hat \psi_{n}^\intercal z_{n} - \hat \psi_n^0 \label{psp:obj} \\
	&\textrm{s.t.} \quad& By_n^k(\zeta_{n_k}) \leq V(\zeta_{n_k})(b + A_nz_n),\quad k\in K_n,\  \zeta_{n_k}\in \Xi^{n_k}\\
	&             & z_n\in \{0,1\}^{G'+L'+S'} \\
	& 	       & y_n^k(\zeta_{n_k}) \in Y_n^k(\zeta_{n_k}), \quad \forall k\in K_n,\ \zeta_{n_k} \in \Xi^{n_k}.
	\end{eqnarray}
\end{subequations}

To attain an integer solution for the original [MM-SMIP], one could embed the column generation procedure within a branch-and-bound framework, resulting in the branch-and-price algorithm \citep{Barnhart1998}, where the solution obtained from [PMP-LR] %\footnote{Note that [PMP-LR] is just [RPMP-LR] (Equation \ref{RPMP_LR}) with the subset $\widetilde{\mathcal{F}}_n$ being replaced by the full set $\mathcal{F}_n$.} 
would represent the \textit{root node} of the branch-and-bound tree. Specially if an optimal solution of [PMP-LR] consists only integers, then we have solved [PMP] at the \textit{root node}. Otherwise, branching would occur when the [PMP-LR] solution does not satisfy the integrality conditions and the column generation is continued throughout the branch-and-bound tree.
It is observed in our numerical experiment, however, optimal integer solutions are always obtained by solving [PMP-LR] to optimality with the column generation algorithm; and thus the branching process is not necessary. This is so because of the perfect-matrix\footnote{Note that a 0-1 matrix is said to be perfect if the polytope of the associated set packing problem only has integral vertices.} structure \citep{Padberg1974} of the constraint matrix corresponding to each variable $\lambda_n^i$, for $n\in \mathcal{N}, i\in \mathcal {F}_n$ in [PMP-LR]. The similar effect is found and discussed in \citep{Ryan1988442} and \citep{SiPhWo09}, which showed that a fractional solution is possible but very rarely obtained from solving the linear relaxation of Dantzig-Wolfe reformulation (aka [PMP-LR] in our case) with such a structure. (As proved by \cite{Padberg1974}, all total unimodular matrices are perfect, but the reverse is not true, hence the possibility of having none integral solutions.) In this paper, we mainly discuss the algorithm for solving [PMP-LR]. Should non-integer solutions appear, we can always embed the algorithm into a branch-and-bound scheme to obtain an optimal integer solution for all cases.

%Singh has discussed the rarely in practice fractional solutions might be obtained, but it is very rare. Main computational cost is still with the column generation in each branch and bound node as the branching time is negligible. the parallel computing would still expect to significantly reduce the wall time. Our paper will focus on solving the initial linear relaxation of PMP, that is, PMP-LR.

%Instead of using the branch-and-price method, we impose the integrality constraints in the \textit{root node}, reaching a restricted formulation of the primary master problem - [RPMP], and solve it to obtain the integer solutions to the MM-SMIP. Let [RPMP]$^\ast$ be the optimal objective function value of [RPMP] and we would expect [PMP-LR]$^\ast \leq \textrm{[PMP]}^\ast = \textrm{Orig}^\ast \leq \textrm{[RPMP]}^\ast$. We will report in our computational result the relative gap, $\frac{\textrm{[RPMP]}^\ast -\textrm{[PMP-LR]}^\ast}{\textrm{[RPMP]}^\ast}$, for each test instance if it is greater than 0.

\subsection{Cutting Plane Approximation of the Operation Costs at the Lower Level}
%Although the nodal decomposition is enabled by the Dantzig-Wolfe reformulation with the splitting variables,
In this subsection, we focus on how to solve the primary subproblem $\textrm{[PSP}_n\textrm{]}$ with $n\in \mathcal{N}$, which may still be intractable due to the large number of simulated scenario paths in the operational level.
%Note that once an expansion plan is known, the problem is then to determine the optimal unit commitment under individual lower-level scenario.
%such that the objective of recourse function ( i.e., the expected operational costs) is minimized.
Each [PSP$_n$]  is essentially a two-stage mixed integer stochastic problem, in which the expansion decisions $z_n$ for the scenario-tree node $n$ are made in the first stage, and the second-stage UC decisions $y_n$ are determined when the uncertainty in the lower level is realized.
We thus propose to further decompose $\textrm{[PSP}_n\textrm{]}$ based on cutting plane approximation of the UC problems. In this way, we effectively separate the operational problems from the expansion planning problem. The following secondary master problem (referred to as [SMP$_n$]) is formulated as the master problem for $\textrm{[PSP}_n\textrm{]}$,
\begin{subequations}
	\begin{eqnarray} \textrm{[SMP}_n\textrm{]}: \quad
	&\min\limits_{z,\ \theta} \quad&  - \hat \psi_{n}^\intercal z_{n} - \hat \psi_n^0 + \theta \label{cep:eq:theta}\\
	& \textrm{s.t.} & z_n\in \{0,1\}^{G'+L'+S'} \\  \label{eq:smp:initcuts}
	& 		& \rho_n^i z_n + \nu_n^i\theta \geq 0, \quad i = 1,2,\dots,Q.  \label{eq:smp:cuts}
	\end{eqnarray}
\end{subequations}
In \eqref{cep:eq:theta}, the variable $\theta$ is an underestimator of the expected second-stage value function $\mathbb{Q}(z_n)$ with respect to the capacity expansion decisions $z_n$.
%represents the optimal second-stage cost of $z_n$;
Constraint \eqref{eq:smp:cuts} represent $Q$ cutting planes as the outer approximation for the recourse function $\mathbb{Q}(z_n)$.  %At each scenario-tree node $n$, let [PSP$_n$]$^\ast$ denote the optimal objective function value of [PSP$_n$], and [SMP$_n$]$^\ast$ be the optimal objective function value of [SMP$_n$].
While the number of cutting planes in \eqref{eq:smp:cuts} can be extremely large,
%it is neither practical nor necessary to include all of them in the master problem \citep{ZhWaPaGu11}. Instead,
we can only include a subset of them at the beginning, resulting in a relaxed formulation (referred to as $\textrm{[rSMP}_n\textrm{]}$); we then apply the integer L-shaped method (cf. \cite{LaLo93}) to iteratively add cuts until an optimal solution $(\theta^\ast,\  z_n^\ast)$ satisfying $\theta^\ast = \mathbb{Q}(z_n^\ast)$ is found, when we would have [PSP$_n$]$^\ast$ = [SMP$_n$]$^\ast$, where [PSP$_n$]$^\ast$ and [SMP$_n$]$^\ast$ denote the optimal objective function values of [PSP$_n$] and [SMP$_n$], respectively.
Given an upper level expansion decision $\hat z_n$ at a scenario-tree node $ n\in \mathcal{N}$, the value of the corresponding recourse function $\mathbb{Q}(\hat z_n)$ is evaluated by solving multiple independent UC problems as in \eqref{DUC}. Each of these UC subproblems corresponds to a given sample path of lower-level uncertainty in a specific day $k$, that is, a specific  $\zeta_{n_k}^p\in \mathbb{P}^{n_k}$, as shown in the following.
\begin{subequations}
	\begin{eqnarray} \textrm{[SP}(\zeta_{n_k}^p)\textrm{]}: \quad
	&\min\limits_y  \quad&  \Rm{TOC}\left(y_n^k(\zeta_{n_k}^p)\right) \\
	&\textrm{s.t.} \quad& B y_n^k(\zeta^p_{n_k}) \leq V(\zeta^p_{n_k})(b + A_n \hat{z}_n) \label{eq:dual_capacity}\\[5pt]
	&             & y_{n}^k(\zeta_{n_k}^p) \in Y_{n}^k(\zeta_{n_k}^p).  \label{eq:dual_others}
	\end{eqnarray}
\end{subequations}

At every iteration, we solve each subproblem and its linear programming relaxation and add: (1) the Benders feasibility cut \citep{BiLo11} if the linear programming relaxation admits no feasible solution, (2) the integer feasibility cut \citep{LaLo93} if the subproblem admits no feasible solution, (3) the Benders optimality cut \citep{BiLo11} if $\theta$ is less than the objective of the linear relaxation, and (4) the integer optimality cut \citep{LaLo93} if $\theta$ is less than the objective of the subproblem.

\subsection{The Nested Cross Decomposition Algorithm}\label{subsec:nested}
\begin{figure}[htb!]
	\begin{center}
		\includegraphics[width=\textwidth]{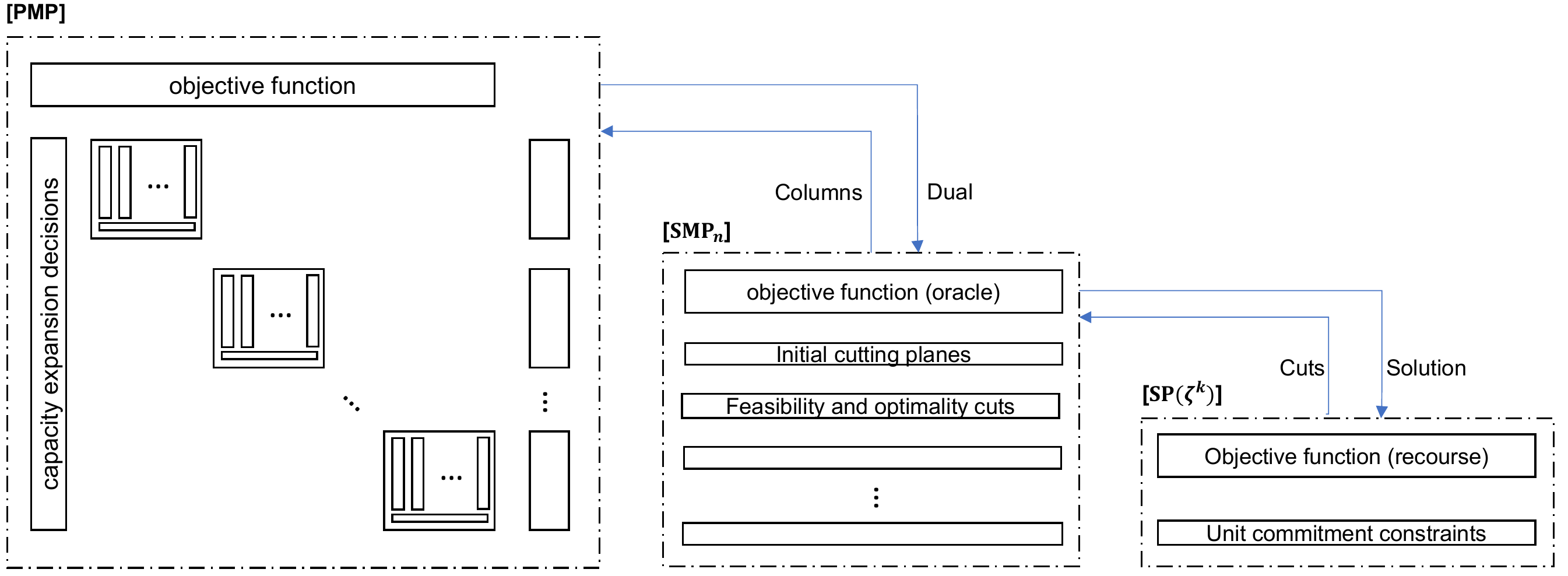}
	\end{center}
	\caption{The nested cross decomposition algorithm}\label{fig:alg}
\end{figure}

In the proposed NCD algorithm, Dantzig-Wolfe decomposition is applied for the upper-level master problem and the integer L-shaped approach is embedded to solve the lower-level UC models. The entire process is shown in Figure \ref{fig:alg}.
Specifically, we apply column generation to solve the upper-level decomposition, [PMP], in which the columns are generated by solving a set of subproblems at each scenario-tree node $n$. Each column represents a capacity expansion plan along the scenario path from the root node to node $n$. Each of these subproblems in the strategic level is independent to each other, and is further decomposed to a secondary master problem, [SMP$_n$], and a list of subproblems, [SP$(\zeta_{n_k}^p)$]'s, in the operational level. The secondary master problem is first relaxed and solved, and the feasibility and optimality condition of its optimal solution is  then checked. The feasibility and optimality cuts are then generated and added to \textrm{[SMP}$_n$\textrm{]} should the condition be violated.

An important feature of our algorithm is that the cutting planes \eqref{eq:smp:cuts} can be carried over to the [SMP$_n$] in next column generation iterations.
It means that the [SMP$_n$] can start with more constraints as the column generation proceeds. This is because the cuts generated in previous iterations are all valid to the secondary master problems afterwards, as presented in Theorem \ref{thm:valid_opt_cuts}.
\begin{theorem} \label{thm:valid_opt_cuts}
For each scenario-tree node $n\in \mathcal N$, the cutting planes \eqref{eq:smp:cuts} that have been generated in previous column generation iterations are valid to the [SMP$_n$] in a later iteration.
\end{theorem}
\begin{proof}
The feasible region of each [SP($\zeta_{n_k}^p$)] dual problem does not depend on the upper level expansion decision $\hat z_n$. A Benders cut that has been generated based on the extreme points of such feasible region, that is obtained by solving [SP($\zeta_{n_k}^p$)]'s in previous column generation iterations when different expansion decisions are given, are thus valid in later iterations.
Meanwhile, the recourse function $\mathbb{Q}(z_n)$ is solely dependent on the expansion decision $z_n$. This means that once the expansion decision is fixed as $\hat z_n$, the resulting recourse function value $\mathbb{Q}(\hat z_n)$ would be determined no matter regardless of column generation iterations. The validity of adding integer L shaped cuts from previous iterations is thus proved.
%\Halmos
\end{proof}
We leverage Theorem \ref{thm:valid_opt_cuts} to integrate the column generation and integer L shaped algorithms in a more efficient way for solving the proposed problem, rather than simply embed one of the two well-known decomposition methods in another.
The proposed NCD algorithm is described in Algorithm \ref{cep:alg:nd}, and we demonstrate in Theorem \ref{thm:convergence} that the algorithm can provide a finite exact approach for solving the [PMP-LR]
In addition, we show in Remark \ref{remark:obj} (Appendix \ref{appendix:remark}) the relationship among the optimal objective function values of the various optimization problems defined within the algorithm.
%This linear programming relaxation to the master problem for Dantzig-Wolfe decomposition would expect to provide a very good approximation for the MM-SMIP.

\begin{theorem} \label{thm:convergence}
The NCD algorithm can find the optimal solution of [PMP-LR] within a finite number of iterations.
\end{theorem}
\proof{Proof.}
First, consider the inner loop of solving [SMP$_n$].
According to \cite{LaLo93}, the integer L-shaped method can provide a finite exact algorithm for the stochastic mixed integer problem [PSP$_n$], in the presence of binary variables in both stages.
Now consider the outer loop of solving [PMP-LR]. The studies in \cite{Dantzig1960} and \cite{SiPhWo09} show that the column generation algorithm can solve the Dantzig-Wolfe reformulation enabled by the nodal decomposition ([PMP-LR] in our case) to optimality after a finite number of steps. The combination of the convergence results are applicable here, and hence our proposed algorithm converges within a finite number of iterations.
\endproof

\begin{algorithm}[htb!]
	\small{\caption{The nested cross decomposition algorithm for the multistage and multiscale stochastic electricity infrastructure investment problem}
	\label{cep:alg:nd}
	\begin{algorithmic}[1]
			\State {\textbf{initialize} the [PMP-LR] with a subset of columns, yielding the restricted problem [RPMP-LR]. }
           \State {initialize [SMP$_n$] with constraints \eqref{eq:smp:initcuts}, yielding the relaxed problem [rSMP$_n$], $\forall n\in \mathcal N$. }
			\Repeat
				\Comment {outer loop of solving [PMP-LR]}
				\State {solve the [RPMP-LR] and extract its dual solution.}
				\ForAll   {$n\in \mathcal{N}$}
					\While {true}
						\Comment {inner loop of solving [SMP$_n$]}
						\State {solve the [rSMP$_n$] and extract the solution $(\hat \theta, \hat z_n)$.}
						\State {solve [SP$(\zeta^p_{n_k})$] linear relaxation, $\forall k\in K_n,\ \zeta^p_{n_k}\in \mathbb{P}^{n_k}$.} \Comment{independent and parallelizable}
						\If {$\hat \theta < \mathbb{Q}_{LP}(\hat z_n)$}
							\State {add Benders cuts to [rSMP$_n$].}
   		                \State {\textbf{continue}}
                       \EndIf
						\State {solve [SP$(\zeta^p_{n_k})$], $\forall k\in K_n,\ \zeta^p_{n_k}\in \mathbb{P}^{n_k}$.} \Comment{independent and parallelizable}
						\If {$\hat \theta \geq \mathbb{Q}(\hat z_n)$}
                           \State {\textbf{break}}									
						\EndIf
                       \State {add integer L shaped cuts to [rSMP$_n$].}		
					\EndWhile				
				\EndFor
				\State {generate and add new columns to [RPMP-LR].}
			\Until {no negative reduced cost column exists, i.e., [SMP$_n$]$^\ast \geq 0, \ \forall n\in \mathcal{N}.$ }
			\State {The [PMP] and thus [MM-SMIP] is solved should the [PMP-LR] optimal solution be integer.}
	\end{algorithmic}
	}
\end{algorithm}
As previously illustrated, we would almost always obtain an integer optimal solution after solving the [PMP-LR] because of the structure of its constraints.
This is further confirmed by our computational experiment, as shown in the next section, in which we have obtained integer solutions for all the test instances by solving this continuous relaxation with the NCD algorithm.

%------------------------------------------------------------------------%
\section{Numerical Experiments and Results}\label{sec:exp}
We test our proposed MM-SMIP model and solution approach in solving instances with various problem scales. The input data, experimental setup and computational results of our numerical experiments are presented in this section. For the illustration purpose, we first test a 6-bus system to demonstrate the benefit of applying the proposed multiscale and multistage stochastic programming approach in the long-term planning problem of electricity infrastructure expansion. The IEEE 118-bus system is then studied to evaluate the NCD algorithm in addressing the intractability of large-scale problems. The algorithm is programmed and compiled with C++ MPI codes, and is run in a HPC cluster with around 3,500 cores, 7.5 TB of RAM, and 56Gb Infiniband interconnect between all nodes. Each node is a shared memory system consisting of 28 cores sharing the same memory resource. The commercial solver, GUROBI, is called for solving all the master problems and subproblems after the decomposition in the algorithm.
%{\color{red} All the input data and programming codes are shared in BitBucket(https://huangzhouchun@bitbucket.org/huangzhouchun/cep.git).}

\subsection{Multiscale Uncertain Parameters Settings} \label{subsec:uncertainty}
%The proposed multiscale model captures coarse-temporal-scale uncertainties, such as investment costs and long-run demand growth rates, through a scenario tree, and fine-temporal-scale uncertainties, such as the hourly electricity demand and penetration of renewable energy. Accordingly, the long-term investment decisions are made in the here-and-now process, and detailed short-term UC problems for some representative days  are included in each node of the scenario tree to ensure that the system after expansions would meet real-time electricity demand in the presence of the uncertainties. 
%The setting regarding these uncertain parameters in our experiments are described in the following.
%\subsubsection{The coarse-temporal-scale uncertainties}

As shown in Figure \ref{fig:scenario_tree}, a strategic-level scenario tree describing the expansion decision making process spans a few investment periods.
Each investment period (or stage) spans five years, and the investment decisions of building new generators of different types, connecting transmission lines and installing energy storage facilities are made in each stage when the long-term uncertain parameters are resolved. The average capital costs of each type of infrastructure in our experiments are presented in Table \ref{table:capital-costs} based on the cost estimates in \cite{CPC2020,GORMAN2019110994} and \cite{Schoenung2011}. We consider that there are 2 possible realizations of investment costs and long-run demand changes at each stage.
In each problem instance the annual peak load at each bus is known at the beginning of the planning horizon, i.e., at the root of the decision tree.
In one realization the costs are 5\% higher than the average costs and the electricity demand grows by 15\%. In contrast, the costs are 5\% lower than the averages and the demand increases by 5\% in the other. %We randomly generate the exact percentage differences for each realization of the large-scale uncertainties.

\begin{table}[!htbp]
  \centering
  \small{
  \caption{The average capital costs of various infrastructure types}
    \begin{tabular}{lcccccr}
    \toprule
    \multirow{2}[0]{*}{Infrastructure}  & \multicolumn{4}{c}{Generator} & \multirow{2}[0]{*}{Transmission Line} & \multirow{2}[0]{*}{Energy Storage} \\ \cline{2-5}
                                        & Coal & Natural gas & Solar Farm & Wind &  & \\ \midrule
    Capital Cost (\$$\slash$kW) & 900 & 900 & 1500 & 1500 & 50 & 400 \\
      \bottomrule
    \end{tabular}
    \label{table:capital-costs}
}
\end{table}

%\begin{table}[!htbp]
%  \centering
%  \small{
%  \caption{The average capital costs of various infrastructure types}
%    \begin{tabular}{r|c}
%    \toprule
%    Infrastructure & \multicolumns{4}{c}{Generator} & \multirow{2}[0]{*}{Transmission Line} & \multirow{2}[0]{*}{Energy Storage}
%
%      Infrastructure & Capital Cost (k\$$\slash$ MW) \\ \hline
%      %\hline
%      % after \\: \hline or \cline{col1-col2} \cline{col3-col4} ...
%      Coal Generator & 900 \\
%      Natural Gas Generator & 900 \\
%      Solar Farm & 1500 \\
%      Wind Generator & 1500 \\
%      Transmission Line & 150 \\
%      Energy Storage & 200 \\
%      \bottomrule
%    \end{tabular}}
%    \label{table:capital_costs}
%\end{table}

%\subsubsection{The fine-temporal-scale uncertainties}\label{subsubsec:finer_uncertainty}
Within each scenario-tree node of a coarse-temporal-scale uncertainty realization, we consider \emph{eight} representative days as the operating periods between investment periods, resulting in eight stochastic UC problems for measuring the reliability and economic impact of investment decisions.
Specifically, we consider four seasons, i.e., spring, summer, fall and winter, to respect that each of them would have significantly different weather conditions (thus different renewable energy penetration) and electricity demand. Weekend days tend to have different electricity usage patterns, as the total demand is usually much lower than on a weekday. Therefore, we consider a weekday and a weekend as the representatives in each season for modeling the UC problems. %For the operation in each representative day, the uncertainty involves the volatility of electricity demand and renewable energy plants' supply. 
We assume that the uncertain parameters in the fine temporal scale, including the hour-to-hour electricity demand and solar/wind power outputs, are unknown except for their probability distributions. More details of the uncertain parameters settings are described in Appendix \ref{appendix:uncertainty}. The Monte Carlo approach is applied to randomly generate a certain number of scenarios for each representative day, which are feed into the multiscale model to ensure that the strategic-level investment decisions render the power system sufficient capacity to satisfy the demand under each scenario.

%\subsection{Computational Results}
\subsection{The Benefit of Considering multiscale Uncertainties}
A 6-bus power system is studied to illustrate the benefit of considering multiscale uncertainties in the power system capacity expansion. The system is constructed according to the original data in \cite{WuZeZhZh2016}. It initially includes two generators, four transmission lines, and three loads. The generating capacity of the system can be expanded by building more generators, for which there are two fossil fuel generators, one solar farm and one wind generator to choose from. In addition, three more transmission lines can be constructed to connect more sites, and two energy storage devices can be installed along with renewable energy power plants. The detailed data for generators, lines, energy storage and load are described in Tables \ref{tab:6bus-gens}-\ref{tab:6bus-load} (Appendix \ref{appendix:6-bus}).

We consider in the case that the upper-level investment decisions are made over three stages, resulting in four scenarios of coarse-temporal-scale uncertainties, and the lower-level operational decisions for each representative day are made with respect to 100 scenarios of fine-temporal-scale uncertainties.
%The optimal solution of capacity expansion obtained from our proposed model is presented in Figure \ref{fig:stochastic_vs_determinstic} Left.
As a benchmark, we consider in the lower level only one deterministic UC problem for each upper-level scenario-tree node by using the expected values of fine-temporal-scale uncertain parameters, which would yield the equivalent model with \cite{SiPhWo09} and \cite{LiSiCo2018}.
%and the solution is shown in Figure \ref{fig:stochastic_vs_determinstic} Right.
The optimal solution of capacity expansion obtained from our proposed model and and that from the benchmark are presented in Table \ref{tab:stochastic_vs_determinstic}.

%The result shows that the use of our multiscale model suggests more capacity expansions with higher investment cost than the benchmark.
%This indicates that our proposed model can render the better system reliability
%However, the solution obtained from the latter is infeasible in the other that considers the uncertainty of hour-to-hour electricity demand and renewable energy penetration.If we we integrate a determinstic lower-level UC in the model, the investment decision obtained is INFEASIBLE in the model that considers stochasticity. Actually, the multiscale model suggests more expansion with higher investment cost, thus rendering better system reliability.
\begin{table}[!htbp]
  \centering \small
  \caption{The investment decisions with and without considering multiscale uncertainties}
  \resizebox{\linewidth}{!}{
    \begin{tabular}{lccccccr}
    \toprule
    \multirow{2}[4]{*}{Scenario} & \multicolumn{3}{c}{Expansion with multiscale uncertainties} & \multicolumn{3}{c}{Expansion without multiscale uncertainties} \\
\cmidrule{2-7}          & Stage 1 & Stage 2 & Stage 3 & Stage 1 & Stage 2 & Stage 3 \\
    \midrule
    1     & Wind, S2,  L2, L5 & G3    & -     & Wind, S2,  L2, L5 & -     & - \\
    2     & Wind, S2,  L2, L5 & G3    & Solar, S1 & Wind, S2,  L2, L5 & -     & G3 \\
    3     & Wind, S2, L2, L5 & G4    & -     & Wind, S2, L2, L5 & G3    & - \\
    4     & Wind, S2, L2, L5 & G4    & Solar, S1, L7 & Wind, S2, L2, L5 & G3    & Solar, S1 \\
    \bottomrule
    \end{tabular}}
  \label{tab:stochastic_vs_determinstic}%
\end{table}%

The result indicates that more expansions of infrastructure are necessary so as to reliably serve future electricity demand when the uncertainty of hour-to-hour electricity demand and renewable energy penetration are considered in the lower-level operations. It is found in our experiment that the system capacity given by the benchmark solution fails to satisfy the demand under many of the lower-level scenarios, and the system is exposed to a high risk in the presence of operational-level random parameters. This verifies the impact of fine-scale uncertainties on long-term planning and the benefit of considering the uncertainties in different time scales to more accurately study the infrastructure needs and maintain reliability of an electricity grid.

%\subsubsection{Sensitivity of decisions to lower-level scenario generations}
% \begin{figure}
%   \centering
%   % Requires \usepackage{graphicx}
%   \includegraphics[width=5in]{figures/sensitivity}\\
%   \caption{The sensitivity of investment and operational costs to lower-level scenario generations}
%   \label{fig:sensitivity}
% \end{figure}

%---------------------------------------------------------------------------------------------------------------------------------------------------------------------------------------------------%

\subsection{Acceleration with Warm Start}
We further analyze the convergence of our multistage and multiscale stochastic programming procedure with the warm start of pricing subproblems, that is, keeping the cutting planes generated in the inner loop of solving each [SMP$_n$] through the outer loop of the NCD algorithm.
\begin{figure}[!htbp]
  \centering
  \includegraphics[width=\textwidth]{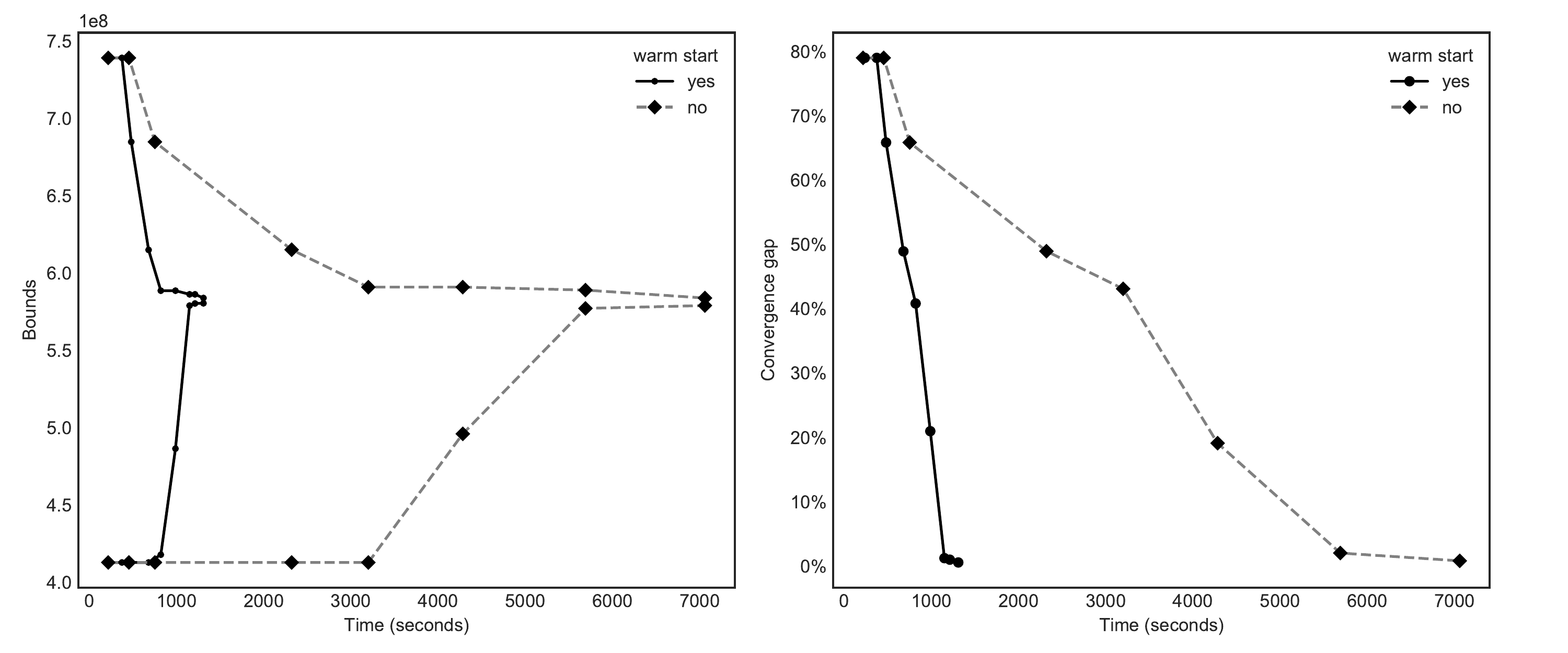}\\
  \caption{The objective-function upper \& lower bounds and convergence gap throughout the NCD algorithm in solvinng the 6-bus power system capacity expansion problem}\label{fig:converge}
\end{figure}
Figure \ref{fig:converge} shows the upper and lower bounds as well as the convergence gap throughout the NCD algorithm in solving the instance in the 6-bus system as described above, with and without the warm start. We employ 28 cores for both of the procedures. 

It can be seen that the warm start strategy significantly reduces the time of convergence.
%it saves 80\% computational time for the algorithm to satisfy the tolerance of 1\%. 
With the strategy, the convergence gap starts to drop very fast and satisfy the tolerance of 1\% within around 20 minutes. When the warm start is not applied, in contrast, the optimality gap is still more than 50\% within the same amount of time. It takes a large amount of time per iteration to solve the pricing problems from scratch. This suggests that many of the cutting planes generated in the previous iterations can help to tighten the feasible region of the [SMP$_n$] for each scenario-tree node $n\in \mathcal N$ in subsequent iterations, and thus improve the algorithm performance. %Similar convergence results are also observed in our computational experiments in all the other test instances.

%---------------------------------------------------------------------------------------------------------------------------------------------------------------------------------------------------%
\subsection{Computational Performance}
A modified IEEE 118-Bus System is studied to demonstrate the practical use of the proposed model in a large power system capacity expansion problem. The system is constructed according to the original data in \cite{IIT2003}. It initially consists of 34 generators, 166 transmission lines and 91 loads. The total installed generating capacity is 4060 MW.
The candidate facilities that can be built over the planning horizon include 20 generators (8 wind generators, 2 solar farms and 10 fossil fuel generators, with the total generating capacity of 3430 MW), 20 transmission lines and 10 energy storage devices. More detailed data of the system are provided in Appendix \ref{appendix:118-bus}.
In such system, we investigate the computational performance of various algorithms by testing a set of problem instances with different upper-level scenario-tree statistics and different simulations of lower-level scenario paths. 
The characteristics of the instances are listed in Table \ref{tab:118-bus-instances}, in which the last three columns present the numbers of variables and constraints in the extensive formulation with respect to each instance.
\begin{table}[!htbp]
  \centering
  \caption{The characteristics of test instances in the IEEE 118-bus system} \small
  \resizebox{\textwidth}{!}{
    \begin{tabular}{lcccrrr}
    \toprule
    \multirow{2}[2]{*}{Instance} & \multirow{2}[2]{*}{\# Stages} & \multicolumn{1}{c}{\# Upper-level} & \multicolumn{1}{c}{\# Lower-level} & \multicolumn{1}{c}{\# Discrete } & \multicolumn{1}{c}{\# Continuous } & \multicolumn{1}{c}{\multirow{2}[2]{*}{\# Constraints}} \\
          &       & \multicolumn{1}{c}{scenarios} & \multicolumn{1}{c}{scenarios} & \multicolumn{1}{c}{variables} & \multicolumn{1}{c}{variables} &  \\
    \midrule
    1     & 3     & 4     & 20    & 2,526,965 & 8,467,200 & 16,719,605 \\
    2     & 4     & 8     & 20    & 5,414,925 & 18,144,000 & 35,827,725 \\
    3     & 5     & 16    & 20    & 11,190,845 & 37,497,600 & 74,043,965 \\
    4     & 6     & 32    & 20    & 22,742,685 & 76,204,800 & 150,476,445 \\
    % 5     & 5     & 81    & 30    & 65,518,475 & 219,542,400 & 433,513,355 \\
    % 6     & 6     & 243   & 30    & 197,096,900 & 660,441,600 & 1,304,122,820 \\
    \midrule
    5     & 3     & 4     & 100   & 12,633,845 & 42,336,000 & 83,597,045 \\
    6     & 4     & 8     & 100   & 27,072,525 & 90,720,000 & 179,136,525 \\
    7     & 5     & 16    & 100   & 55,949,885 & 187,488,000 & 370,215,485 \\
    8    & 6     & 32    & 100   & 113,704,605 & 381,024,000 & 752,373,405 \\
    % 11    & 5     & 81    & 100   & 218,385,035 & 731,808,000 & 1,445,034,635 \\
    % 12    & 6     & 243   & 100   & 656,959,940 & 2,201,472,000 & 4,347,046,340 \\
    \bottomrule
    \end{tabular}}
  \label{tab:118-bus-instances}%
\end{table}

We investigate the computational performance of the NCD algorithm in comparison against (1)Direct: directly solving the extensive formulation of each instance with the commercial solver, i.e., GUROBI 9.0.2, and (2) CG: column generation algorithm which solves the pricing subproblems $\textrm{[PSP}_n, \forall n \in \mathcal{N}\textrm{]}$ directly without decomposition, and 3)Benders decomposition: the Benders decomposition algorithm with all the investment decision variables and constraints included in a master problem and the operational variables and constraints in subproblems. The results, including the solution time (in seconds) and the best objective value obtained from each algorithm, are reported in Table \ref{tab:118-bus-results}. All the methods employ the same computer resource for solving each instance. The numbers of cores are displayed in the column ``\#Cores". However, the ``Direct" method can only call one computer node, which contains 28 cores, to solve the extensive model of each procedure.
Each program is terminated either when the optimality gap of 1\% is reached or the time limit of 48 hours is reached. 
%Except for the ``Direct" method that can only calls one computer node for each procedure, all the other methods employ exactly the same computer resources. 
%The results are reported in Table \ref{tab:118-bus-results}. Note that each program for each instance is terminated when the time limit of 48 hours is reached, and the optimality gap is reported.
\begin{table} 
  \caption{Comparison of computational performance}
  \resizebox{\linewidth}{!}{
    \begin{tabular}{lccrrrr}
      \toprule
      \multirow{2}[4]{*}{Instance} & \multirow{2}[4]{*}{\# Cores} & \multirow{2}[4]{*}{Direct/CG} & \multicolumn{2}{c}{Benders decomposition} & \multicolumn{2}{c}{Nested cross decomposition} \\
  \cline{4-7}          &       &       & Time (seconds) & Best obj.(million) & Time (seconds) & Best obj.(million) \\
      \midrule
      1     & 28    & $-$     & 672   & 4957.60 & 4 860  & 4945.75 \\
      2     & 60    & $-$     & 1 795  & 6802.91 & 13 970 & 6799.41 \\
      3     & 124   &$-$      & 10 225 & 8567.70 & 31 235 & 8555.87 \\
      4     & 252   & $-$     & 57 440 &  10310.00     & 55 410 & 10295.30 \\
      \midrule
      5     & 28    & $-$     & 3 269  & 4971.35 & 28 360 & 4969.27 \\
      6     & 60    & $-$     & 157 420 & 6825.71 & 167 900 & 6814.15 \\
      7     & 124   & $-$     & $-$ & $-$ & 2.67\% & 8624.17 \\ %\down
      8     & 252   & $-$     & $-$ & $-$ & 5.31\% & 10492.30 \\
      \bottomrule \vspace{-12pt}
      \end{tabular}} 
      {\emph{Notes.} A percentage represents the optimality gap obtained when the time limit of 48 hours is reached. A dash indicates that no solution has been found.}
  \label{tab:118-bus-results}%
\end{table}%

It is observed in our experiments that for every problem instance the program runs out of memory in the cluster due to the extremely large scale when its extensive formulation is directly managed. The standard column generation fails to find a feasible solution within the time limit for each instance either. 
The Benders decomposition algorithm exhibits the better computational performance, especially for small instances in which it even outperforms the NCD approach. However, it has difficulty in finding a feasible solution for large instances with more than 4 planning stages and 100 lower-level scenarios. This is because the Benders feasibility cuts cannot eliminate infeasible investment decisions efficiently enough when the capacity expansion decision variables are all included in the algorithm's master problem. It also spend significantly more time per iteration when the algorithm proceeds and more cuts are added into the single master problem.
In contrast, the NCD algorithm exhibits the better efficiency for large problem instances. On one hand, the computational difficulty of pricing subproblems in the column generation is addressed by the second-layer L-shaped decomposition. On the other hand, with the first-layer Dantzig-Wolfe decomposition, each iteration of the column generation is to find a set of feasible capacity expansion solutions with respect to each scenario-tree node, and this is easier than seeking the feasible solution for the whole scenario tree in the Benders decomposition approach.  

\begin{table}[htbp]
  \centering
  \caption{Computational results of NCD algorithm}
  \begin{tabular}{lccrrrr}
    \toprule
    \multirow{2}[4]{*}{Instance} & \multicolumn{2}{c}{\# Iterations} & \multicolumn{2}{c}{Time (seconds)} & \multicolumn{2}{c}{Objective (million)} \\
\cmidrule{2-7}          & 5\%   & 1\%   & 5\%   & 1\%   & 5\%   & \multicolumn{1}{c}{1\%} \\
    \midrule
    1     & 16    & 21    & 4 130  & 4 860  & 5063.27 & 4945.75 \\
    2     & 20    & 39    & 7 155  & 13 970 & 6865.64 & 6799.41 \\
    3     & 19    & 53    & 14 350 & 31 235 & 8669.57 & 8555.87 \\
    4     & 13    & 35    & 25 450 & 55 410 & 10458.20 & 10295.30 \\ \hline
    5     & 13    & 22    & 12 290 & 28 360 & 5002.12 & 4969.27 \\
    6     & 19    & 32    & 65 460 & 167 900 & 6879.00 & 6814.15 \\
    7     & 20    & 46    & 101 900 & 254 400 & 8705.78 & 8558.17 \\
    8     & 13    & 38      & 213 300 & 576 400 & 10477.60 & 10305.40 \\
    \bottomrule
    \end{tabular}
  \label{tab:ncg_result}%
\end{table}%

As a final test, we remove the time limit of the NCD algorithm in order to investigate the convergence results especially for large instances.
Table \ref{tab:ncg_result} displays the number of major iterations (in the outer loop) and the solution time for the algorithm to reach the gaps of 5\% and 1\%, respectively, along with the corresponding objective values. In fact, the Benders decomposition algorithm is still not able to find a feasible solution within the same amount of time as in the NCD algorithm for the largest two instances (instances 7 and 8). It is important to note that there is almost no difference in the numbers of iterations between the instances 1-4 and 5-8. However, it takes a lot more solution time to solve the latter instances compared to the former. This is because the embedded L-shaped method experiences more computational difficulty of handling pricing subproblems when more scenario paths are considered in the operational level. It is also interesting to observe in our experiments that the upper bound of each problem is reduced very quickly toward the optimal value through early iterations of the NCD algorithm. After each procedure reaches the 5\% gap, the upper bound drops very slightly, and it is the increase of lower bound that mainly contributes to the convergence toward the 1\% gap. For this reason, there is no significant difference between the objective value with the optimality gap of 5\% and 1\% for each instance.    

\section{Conclusions}\label{sec:con}
In this paper, we propose a multiscale, multistage stochastic mixed integer programming model for power systems expansion planning. This model takes into consideration not only the long-term expansion planning of infrastructures including generators, transmission line segments and storage facilities, but also the detailed system operations based on solving unit commitment models. The model captures multiple time scales (years, days and hours) and various sources of uncertainties (investment cost, intermittent renewable energy sources, electric demand etc.), which makes the proposed problem extremely large-scale and computationally very challenging. We design a nested cross decomposition algorithm, which solves the multistage stochastic expansion planning part using column generation and separates unit commitment problems from the expansion planning by cutting plane approximations of the stochastic generation cost functions. We implement the nested cross decomposition algorithm in a parallel computing setting, as there are a gigantic amount of independent subproblems. The effectiveness and efficiency of the approach are demonstrated by our numerical experiments. Our approach provides a possible way to combine long-term planning with operational details and an efficient algorithm to solve such large-scale instances. 

Future efforts in this area could (i) use the proposed model and algorithm to analyze various capacity expansion plans in a real-world power system with relatively large and complex network; and (ii) extend the problem to incorporate risk management, e.g., by modeling chance constraints in the operational level. 

\bibliographystyle{informs2014}
\bibliography{references, literature} % if more than one, comma separated

\clearpage
\appendixpage
\addappheadtotoc
\appendix

\section{Appendix to Nested Cross Decomposition Algorithm (Section \ref{subsec:nested})} \label{appendix:remark}
\begin{remark} \label{remark:obj}
In the inner loop of solving each [SMP$_n$], we have [rSMP$_n$]$^\ast \leq $
[SMP$_n$]$^\ast$ at each iteration (since [rSMP$_n$] is the relaxed version of [SMP$_n$]; i.e., with fewer cutting planes) before the convergence of the integer L-shaped method, when we would have [rSMP$_n$]$^\ast = $  [SMP$_n$]$^\ast =$
[PSP$_n$]$^\ast$. In the outer loop of solving [PMP-LR], [RPMP-LR]$^\ast$ provides an upper bound to the [PMP-LR]$^\ast$ at each iteration; that is, [RPMP-LR]$^\ast \geq $ [PMP-LR]$^\ast$ (since [RPMP-LR] is a restricted version of [PMP-LR]), until the column generation algorithm converges, when we would have  [RPMP-LR]$^\ast =$ [PMP-LR]$^\ast$. Finally, if the original binary variables before being continuously relaxed in [PMP-LR]$^\ast$ all assume integer values in an optimal solution of  [PMP-LR], then we have the corresponding [PMP-LR]$^\ast$ = [PMP]$^\ast$ =  [MM-SMIP]$^\ast$; otherwise, as mentioned earlier in Section \ref{subsec:cg}, a branch-and-bound type of scheme can be employed to recover an optimal solution of [MM-SMIP] from  [PMP-LR]$^\ast$.
\end{remark}

\section{Appendix to Fine-scale Uncertainty (Section \ref{subsec:uncertainty})}\label{appendix:uncertainty}
\begin{figure}[!htbp]
  \centering
  \includegraphics[width=3.5in]{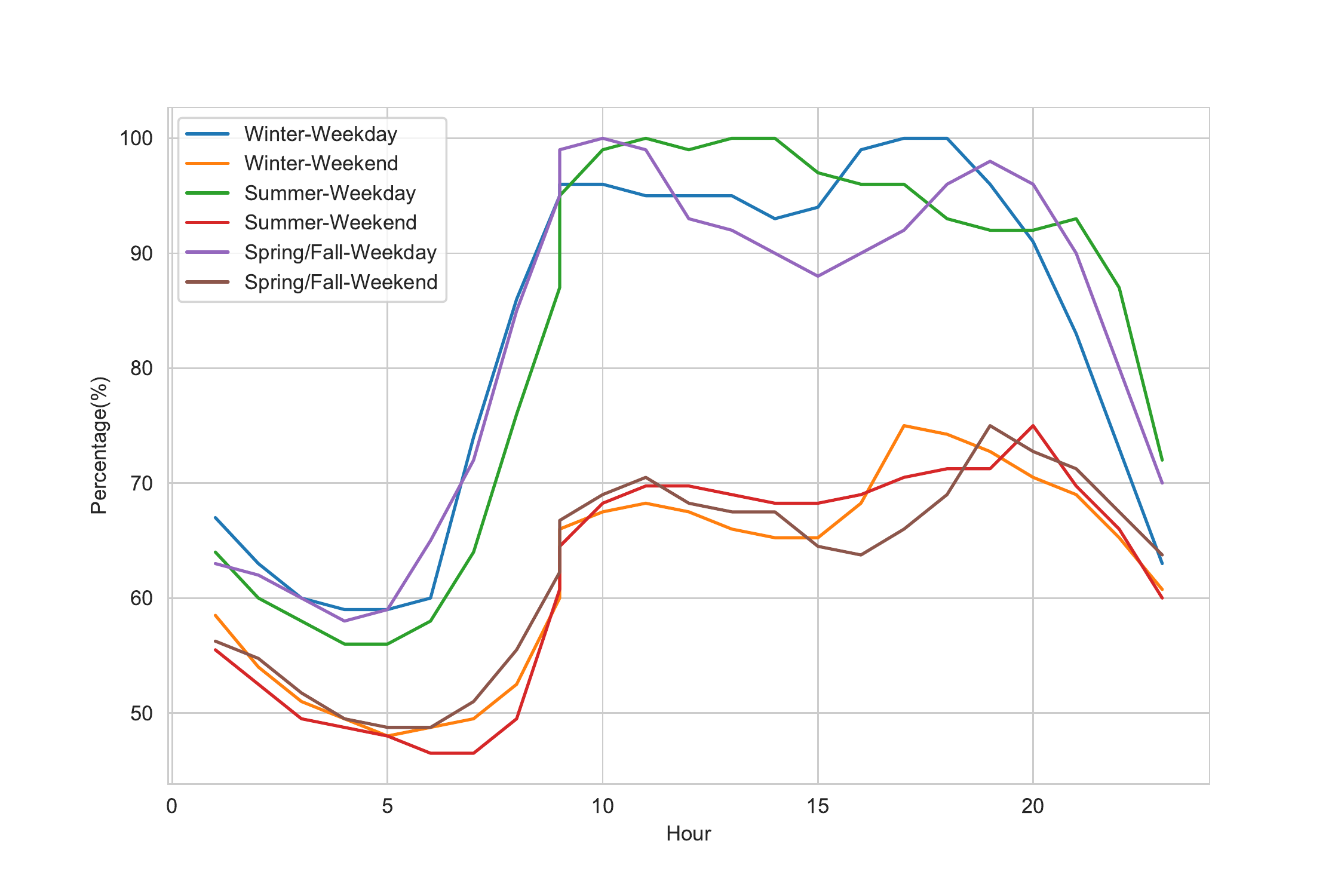}\\
  \caption{The expected electricity demand in percent of annual peak load.}
  \label{fig:expected_demand}
\end{figure}

\begin{figure}[!htbp]
  \centering
  \includegraphics[width=0.45\linewidth]{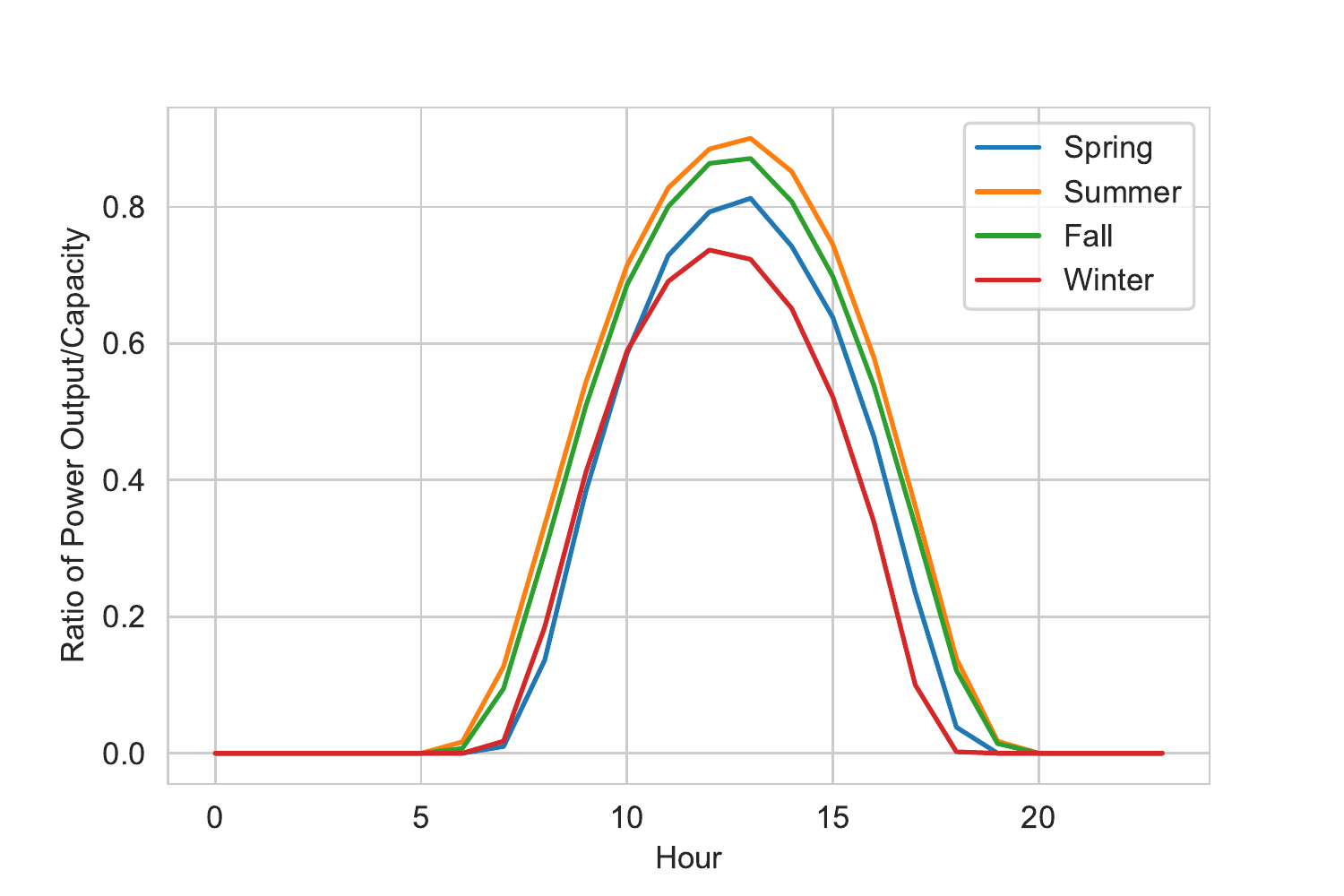}
  \includegraphics[width=0.45\linewidth]{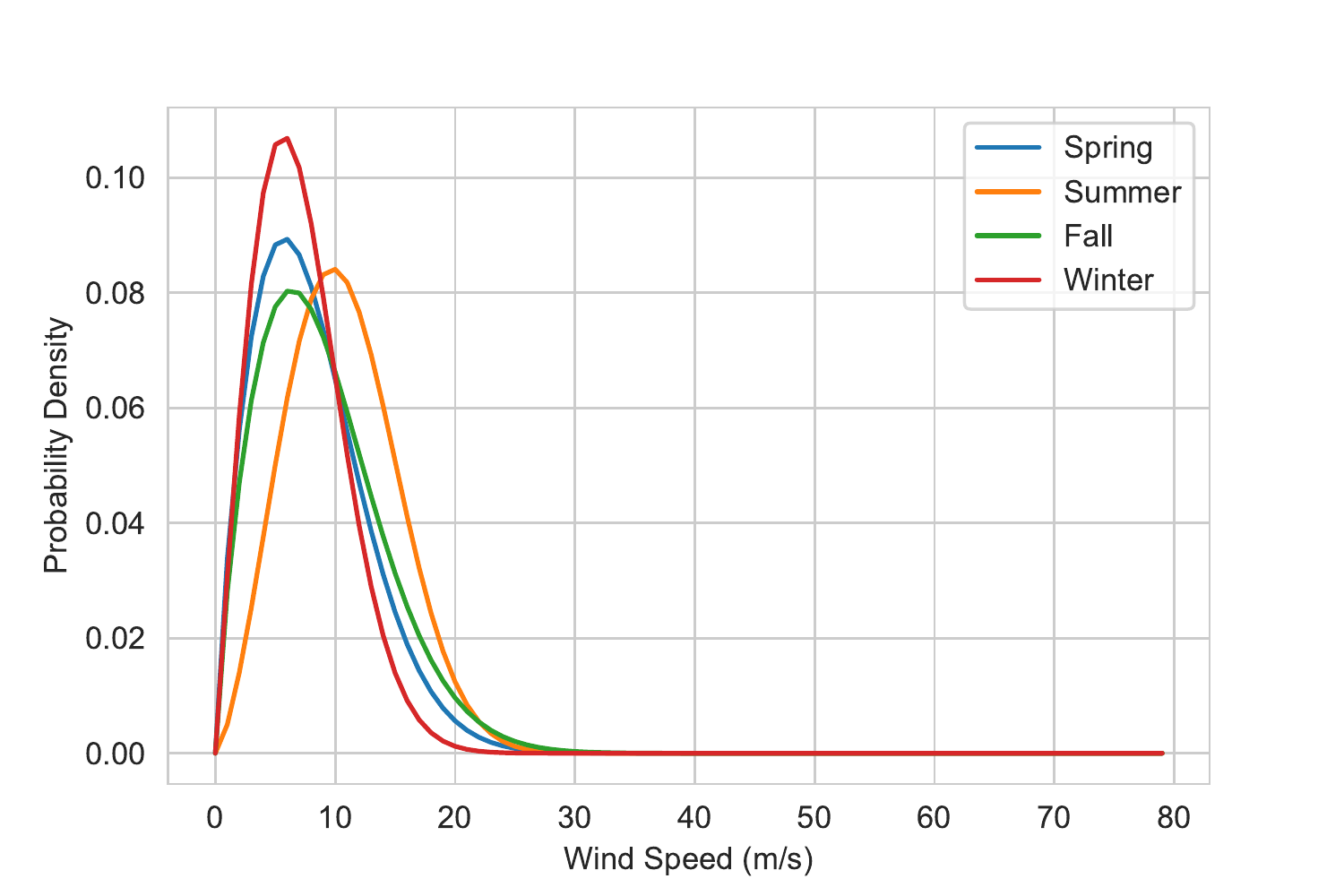}\\
  \caption{Left: The expected ratio of power output to generating capacity of a solar farm within a day in different seasons; Right: Weibull distributions of wind speed values in different seasons.}
  \label{fig:renewable}
\end{figure}

The load profile for each representative day is shown in Figure \ref{fig:expected_demand}. We assume that the electricity demand in each hour follows a normal distribution, of which the mean is given by the profile and the standard deviation is 10\% of the mean.
The renewable energy resources of both solar and wind are considered in our experiments and the intermittency of solar/wind plants' outputs are modeled.
The solar intensity varies among different seasons, yielding various power generations.
Figure \ref{fig:renewable} Left shows the hour-to-hour expected power output compared to the generating capacity of a solar plant within a day for each season. We assume that the energy generated by a solar farm in each hour follows a normal distribution and the standard deviation is 10\% of its mean.
The variability of wind speed also propagates to the power output of wind generators. Weibull distribution is a good fit to the probability distribution of wind speed in a region \citep{SANSAVINI201471}. As shown in Figure \ref{fig:renewable} Right, we consider a Weibull distribution with distinct scale and shape parameters for each season. In our experiments, we assume that a wind generator starts producing power when wind speed equals the cutin speed of $3 m\slash s$, and the power out grows linearly with the wind speed until it reaches the maximum (i.e., the generating capacity of the generator) at the speed of $12 m\slash s$.

\section{Appendix to 6-Bus System Data}\label{appendix:6-bus}
The tested 6-Bus system includes two generators, four transmission lines, and three loads at the beginning of the planning horizon. There are $4$ candidate generators that can be built to expand the current power system, including 2 fossil fuel generators, 1 solar farm and 1 wind generator. In addition, three more transmission lines can be constructed to connect the sites, and two energy storage devices can be installed along with renewable energy power plants.
\begin{table}[!htbp]
  \centering
  \small{
  \caption{Generators in the IEEE 6-Bus system}
  \resizebox{\linewidth}{!}{
    \begin{tabular}{lccccccccccr}
    \toprule
    \multirow{2}[3]{*}{Unit} & \multirow{2}[3]{*}{Bus No.} & Existing & \multicolumn{3}{c}{Generation Cost Coefficient} & Startup Cost & Pmin  & Pmax  & Min On & Min Off & Ramp \\
\cmidrule{4-6}          &       & (Y/N) & a(\$) & b(\$/MW) & c(\$/MW2) &  (\$) & (MW)  & (MW)  & (h)   & (h)   & (MW/h) \\
    \midrule
    G1    & 1     & Y     & 177   & 13.5  & 0.00045 & 100   & 100   & 220   & 4     & 4     & 50 \\
    G2    & 2     & Y     & 130   & 40    & 0.001 & 200   & 10    & 100   & 2     & 3     & 40 \\
    Wind  & 6     & N     & 0     & 0     & 0     & 0     & 0     & 50   & 0     & 0     & 50 \\
    G3    & 1     & N     & 59    & 22.9  & 0.0098 & 45    & 10    & 50    & 1     & 1     & 25 \\
    G4    & 2     & N     & 130   & 32.6  & 0.001 & 300   & 10    & 100   & 2     & 3     & 40 \\
    Solar & 3     & N     & 0     & 0     & 0     & 0     & 0     & 20    & 0     & 0     & 20 \\
 \bottomrule
    \end{tabular}}
  \label{tab:6bus-gens}}
\end{table}%

\begin{table}[!htbp]
  \centering
  \caption{Transmission lines in the IEEE 6-Bus system}
  \small
    \begin{tabular}{lcccr}
    \toprule
    Line No. & From Bus & To Bus & Existing (Y/N) & Flow Limit (MW) \\
    \midrule
    L1     & 1     & 2     & Y     & 200 \\
    L2     & 2     & 3     & N     & 100 \\
    L3     & 1     & 4     & Y     & 150 \\
    L4     & 2     & 4     & Y     & 200 \\
    L5     & 4     & 5     & N     & 200 \\
    L6     & 5     & 6     & Y     & 200 \\
    L7     & 3     & 6     & N     & 100 \\ \bottomrule
    \end{tabular}%
  \label{tab:6bus-lines}
\end{table}%

\begin{table}[!htbp]
  \centering
  \caption{Energy storage in the IEEE 6-Bus system}
  \small
    \begin{tabular}{lccr} \toprule
    Storage No. & Bus No. & Existing (Y/N) & Capacity (MW) \\
    \midrule
    S1     & 3     & N     & 10 \\
    S2     & 6     & N     & 10 \\\bottomrule
    \end{tabular}%
  \label{tab:6bus-storage}%
\end{table}%

\begin{table}[!htbp]\small
  \centering
  \caption{Load data in the IEEE 6-Bus system}
    \begin{tabular}{lc} \toprule
    Bus No. & Annual Peak Load (MW) \\
    \midrule
    3     & 50 \\
    4     & 120 \\
    5     & 120 \\ \bottomrule
    \end{tabular}%
  \label{tab:6bus-load}%
\end{table}

\section{Appendix to Modified IEEE 118-Bus System}\label{appendix:118-bus}
The system is constructed according to the original data in \citep{IIT2003}. It initially consists of 34 generators, 166 transmission lines and 91 loads. The total installed generating capacity is 4060 MW.
The candidate facilities that can be built over the planning horizon include 20 generators (8 wind generators, 2 solar farms and 10 fossil fuel generators, with the total generating capacity of 3430 MW), 20 transmission lines and 10 energy storage devices.

\scriptsize{
\begin{longtable}[c]{lrcccccccccr}
  \caption{Generators in the modified IEEE 118-Bus system\label{tab:118-bus-gen}} \\
    \hline
    \multirow{2}[4]{*}{Unit} & \multirow{2}[4]{*}{Bus No.} & Existing & \multicolumn{3}{c}{Generation Cost} & Starup Cost  & Pmin  & Pmax  & Min On & Min Off & Ramp \\
\cline{4-6}          &       & (Y/N) & a (\$) & b (\$/MWh) & c (\$/MWh\^2) & (\$)  & (MW)  & (MW)  & (h)   & (h)   & (MW/h) \\
    \hline
    \endfirsthead

    \hline
    \endhead

    \hline
    \endfoot

    \hline
    \endlastfoot

    G1    & 4     & Y     & 31.67 & 26.2438 & 0.069663 & 40    & 5     & 30    & 1     & 1     & 15 \\
    G2    & 6     & Y     & 31.67 & 26.2438 & 0.069663 & 40    & 5     & 30    & 1     & 1     & 15 \\
    G3    & 8     & Y     & 31.67 & 26.2438 & 0.069663 & 40    & 5     & 30    & 1     & 1     & 15 \\
    WIND1 & 10    & N     & 0     & 0     & 0     & 0     & 0     & 300   & 0     & 0     & 300 \\
    G4    & 12    & Y     & 6.78  & 12.8875 & 0.010875 & 110   & 100   & 300   & 8     & 8     & 150 \\
    G5    & 15    & Y     & 31.67 & 26.2438 & 0.069663 & 40    & 10    & 30    & 1     & 1     & 15 \\
    G6    & 18    & Y     & 10.15 & 17.82 & 0.0128 & 50    & 25    & 100   & 5     & 5     & 50 \\
    G7    & 19    & Y     & 31.67 & 26.2438 & 0.069663 & 40    & 5     & 30    & 1     & 1     & 15 \\
    WIND2 & 24    & N     & 0     & 0     & 0     & 0     & 0     & 300   & 0     & 0     & 300 \\
    G8    & 25    & Y     & 6.78  & 12.8875 & 0.010875 & 100   & 100   & 300   & 8     & 8     & 150 \\
    G9    & 26    & N     & 32.96 & 10.76 & 0.003 & 100   & 100   & 350   & 8     & 8     & 175 \\
    G10   & 27    & Y     & 31.67 & 26.2438 & 0.069663 & 40    & 8     & 30    & 1     & 1     & 15 \\
    G11   & 31    & N     & 31.67 & 26.2438 & 0.069663 & 40    & 8     & 30    & 1     & 1     & 15 \\
    WIND3 & 32    & N     & 0     & 0     & 0     & 0     & 0     & 100   & 5     & 5     & 100 \\
    G12   & 34    & Y     & 31.67 & 26.2438 & 0.069663 & 40    & 8     & 30    & 1     & 1     & 15 \\
    G13   & 36    & N     & 10.15 & 17.82 & 0.0128 & 50    & 25    & 100   & 5     & 5     & 50 \\
    G14   & 40    & Y     & 31.67 & 26.2438 & 0.069663 & 40    & 8     & 30    & 1     & 1     & 15 \\
    G15   & 42    & Y     & 31.67 & 26.2438 & 0.069663 & 40    & 8     & 30    & 1     & 1     & 15 \\
    WIND4 & 46    & N     & 0     & 0     & 0     & 0     & 0     & 100   & 5     & 5     & 100 \\
    G16   & 49    & Y     & 28    & 12.3299 & 0.002401 & 100   & 50    & 250   & 8     & 8     & 125 \\
    G17   & 54    & N     & 28    & 12.3299 & 0.002401 & 100   & 50    & 250   & 8     & 8     & 125 \\
    G18   & 55    & Y     & 10.15 & 17.82 & 0.0128 & 50    & 25    & 100   & 5     & 5     & 50 \\
    G19   & 56    & N     & 10.15 & 17.82 & 0.0128 & 50    & 25    & 100   & 5     & 5     & 50 \\
    WIND5 & 59    & N     & 0     & 0     & 0     & 0     & 0     & 200   & 8     & 8     & 200 \\
    G20   & 61    & Y     & 39    & 13.29 & 0.0044 & 100   & 50    & 200   & 8     & 8     & 100 \\
    G21   & 62    & N     & 10.15 & 17.82 & 0.0128 & 50    & 25    & 100   & 5     & 5     & 50 \\
    G22   & 65    & Y     & 64.16 & 8.3391 & 0.01059 & 250   & 100   & 420   & 10    & 10    & 210 \\
    G23   & 66    & N     & 64.16 & 8.3391 & 0.01059 & 250   & 100   & 420   & 10    & 10    & 210 \\
    WIND6 & 69    & N     & 0     & 0     & 0     & 0     & 0     & 300   & 0     & 0     & 300 \\
    G24   & 70    & Y     & 74.33 & 15.4708 & 0.045923 & 45    & 30    & 80    & 4     & 4     & 40 \\
    G25   & 72    & N     & 31.67 & 26.2438 & 0.069663 & 40    & 10    & 30    & 1     & 1     & 15 \\
    G26   & 73    & Y     & 31.67 & 26.2438 & 0.069663 & 40    & 5     & 30    & 1     & 1     & 15 \\
    G27   & 74    & Y     & 17.95 & 37.6968 & 0.028302 & 30    & 5     & 20    & 1     & 1     & 10 \\
    G28   & 76    & Y     & 10.15 & 17.82 & 0.0128 & 50    & 25    & 100   & 5     & 5     & 50 \\
    WIND7 & 77    & Y     & 0     & 0     & 0     & 0     & 0     & 100   & 5     & 5     & 100 \\
    G29   & 80    & Y     & 6.78  & 12.8875 & 0.010875 & 440   & 150   & 300   & 8     & 8     & 150 \\
    G30   & 82    & N     & 10.15 & 17.82 & 0.0128 & 50    & 25    & 100   & 5     & 5     & 50 \\
    G31   & 85    & Y     & 31.67 & 26.2438 & 0.069663 & 40    & 10    & 30    & 1     & 1     & 15 \\
    G32   & 87    & Y     & 32.96 & 10.76 & 0.003 & 440   & 100   & 300   & 8     & 8     & 150 \\
    G33   & 89    & Y     & 6.78  & 12.8875 & 0.010875 & 400   & 50    & 200   & 8     & 8     & 100 \\
    G34   & 90    & Y     & 17.95 & 37.6968 & 0.028302 & 30    & 8     & 20    & 1     & 1     & 10 \\
    WIND8 & 91    & Y     & 0     & 0     & 0     & 0     & 0     & 50    & 1     & 1     & 50 \\
    G35   & 92    & Y     & 6.78  & 12.8875 & 0.010875 & 100   & 100   & 300   & 8     & 8     & 150 \\
    G36   & 99    & N     & 6.78  & 12.8875 & 0.010875 & 100   & 100   & 300   & 8     & 8     & 150 \\
    G37   & 100   & Y     & 6.78  & 12.8875 & 0.010875 & 110   & 100   & 300   & 8     & 8     & 150 \\
    G38   & 103   & Y     & 17.95 & 37.6968 & 0.028302 & 30    & 8     & 20    & 1     & 1     & 10 \\
    G39   & 104   & N     & 10.15 & 17.82 & 0.0128 & 50    & 25    & 100   & 5     & 5     & 50 \\
    SOLAR1 & 105   & N     & 0     & 0     & 0     & 0     & 0     & 100   & 0     & 0     & 100 \\
    G40   & 107   & Y     & 17.95 & 37.6968 & 0.028302 & 30    & 8     & 20    & 1     & 1     & 10 \\
    G41   & 110   & N     & 58.81 & 22.9423 & 0.009774 & 45    & 25    & 50    & 2     & 2     & 25 \\
    G42   & 111   & Y     & 10.15 & 17.82 & 0.0128 & 50    & 25    & 100   & 5     & 5     & 50 \\
    SOLAR2 & 112   & N     & 0     & 0     & 0     & 0     & 0     & 100   & 0     & 0     & 100 \\
    G43   & 113   & Y     & 10.15 & 17.82 & 0.0128 & 50    & 25    & 100   & 5     & 5     & 50 \\
    G44   & 116   & Y     & 58.81 & 22.9423 & 0.009774 & 45    & 25    & 50    & 2     & 2     & 25 \\
\end{longtable}}

\scriptsize{
\begin{longtable}{lcccrrcr}
  \caption{Transmission lines in the modified IEEE 118-Bus system\label{tab:118-bus-line}} \\
  \hline
    \multicolumn{1}{l}{From Bus} & \multicolumn{1}{l}{To Bus} & Existing (Y/N) & Flow Limit (MW) & \multicolumn{1}{l}{From Bus} & \multicolumn{1}{l}{To Bus} & Existing (Y/N) & Flow Limit (MW) \\
    \hline
    \endfirsthead

    \hline
    \multicolumn{1}{l}{From Bus} & \multicolumn{1}{l}{To Bus} & Existing (Y/N) & Flow Limit (MW) & \multicolumn{1}{l}{From Bus} & \multicolumn{1}{l}{To Bus} & Existing (Y/N) & Flow Limit (MW) \\
    \hline
    \endhead

    \hline
    \endfoot

    \hline
    \endlastfoot
    1     & 2     & Y     & 175   & 63    & 64    & Y     & 500 \\
    1     & 3     & Y     & 175   & 64    & 61    & Y     & 500 \\
    4     & 5     & Y     & 500   & 38    & 65    & Y     & 500 \\
    3     & 5     & Y     & 175   & 64    & 65    & Y     & 500 \\
    5     & 6     & Y     & 175   & 49    & 66    & Y     & 500 \\
    6     & 7     & Y     & 175   & 49    & 66    & N     & 500 \\
    8     & 9     & Y     & 500   & 62    & 66    & Y     & 175 \\
    8     & 5     & Y     & 500   & 62    & 67    & Y     & 175 \\
    9     & 10    & N     & 500   & 65    & 66    & Y     & 500 \\
    4     & 11    & Y     & 175   & 66    & 67    & Y     & 175 \\
    5     & 11    & Y     & 175   & 65    & 68    & Y     & 500 \\
    11    & 12    & Y     & 175   & 47    & 69    & Y     & 175 \\
    2     & 12    & Y     & 175   & 49    & 69    & Y     & 175 \\
    3     & 12    & Y     & 175   & 68    & 69    & Y     & 500 \\
    7     & 12    & Y     & 175   & 69    & 70    & N     & 500 \\
    11    & 13    & Y     & 175   & 24    & 70    & Y     & 175 \\
    12    & 14    & Y     & 175   & 70    & 71    & Y     & 175 \\
    13    & 15    & N     & 175   & 24    & 72    & Y     & 175 \\
    14    & 15    & Y     & 175   & 71    & 72    & Y     & 175 \\
    12    & 16    & Y     & 175   & 71    & 73    & Y     & 175 \\
    15    & 17    & Y     & 500   & 70    & 74    & Y     & 175 \\
    16    & 17    & Y     & 175   & 70    & 75    & Y     & 175 \\
    17    & 18    & Y     & 175   & 69    & 75    & Y     & 500 \\
    18    & 19    & Y     & 175   & 74    & 75    & N     & 175 \\
    19    & 20    & Y     & 175   & 76    & 77    & Y     & 175 \\
    15    & 19    & Y     & 175   & 69    & 77    & Y     & 175 \\
    20    & 21    & N     & 175   & 75    & 77    & Y     & 175 \\
    21    & 22    & Y     & 175   & 77    & 78    & Y     & 175 \\
    22    & 23    & Y     & 175   & 78    & 79    & Y     & 175 \\
    23    & 24    & Y     & 175   & 77    & 80    & Y     & 500 \\
    23    & 25    & Y     & 500   & 77    & 80    & Y     & 500 \\
    26    & 25    & Y     & 500   & 79    & 80    & Y     & 175 \\
    25    & 27    & Y     & 500   & 68    & 81    & N     & 500 \\
    27    & 28    & Y     & 175   & 81    & 80    & Y     & 500 \\
    28    & 29    & Y     & 175   & 77    & 82    & Y     & 200 \\
    30    & 17    & N     & 500   & 82    & 83    & Y     & 200 \\
    8     & 30    & Y     & 175   & 83    & 84    & Y     & 175 \\
    26    & 30    & Y     & 500   & 83    & 85    & Y     & 175 \\
    17    & 31    & Y     & 175   & 84    & 85    & Y     & 175 \\
    29    & 31    & Y     & 175   & 85    & 86    & Y     & 500 \\
    23    & 32    & Y     & 140   & 86    & 87    & Y     & 500 \\
    31    & 32    & Y     & 175   & 85    & 88    & N     & 175 \\
    27    & 32    & Y     & 175   & 85    & 89    & Y     & 175 \\
    15    & 33    & Y     & 175   & 88    & 89    & Y     & 500 \\
    19    & 34    & N     & 175   & 89    & 90    & Y     & 500 \\
    35    & 36    & Y     & 175   & 89    & 90    & Y     & 500 \\
    35    & 37    & Y     & 175   & 90    & 91    & Y     & 175 \\
    33    & 37    & Y     & 175   & 89    & 92    & Y     & 500 \\
    34    & 36    & Y     & 175   & 89    & 92    & Y     & 500 \\
    34    & 37    & Y     & 500   & 91    & 92    & Y     & 175 \\
    38    & 37    & Y     & 500   & 92    & 93    & N     & 175 \\
    37    & 39    & Y     & 175   & 92    & 94    & Y     & 175 \\
    37    & 40    & Y     & 175   & 93    & 94    & Y     & 175 \\
    30    & 38    & N     & 175   & 94    & 95    & Y     & 175 \\
    39    & 40    & Y     & 175   & 80    & 96    & Y     & 175 \\
    40    & 41    & Y     & 175   & 82    & 96    & Y     & 175 \\
    40    & 42    & Y     & 175   & 94    & 96    & Y     & 175 \\
    41    & 42    & Y     & 175   & 80    & 97    & Y     & 175 \\
    43    & 44    & Y     & 175   & 80    & 98    & Y     & 175 \\
    34    & 43    & Y     & 175   & 80    & 99    & N     & 200 \\
    44    & 45    & Y     & 175   & 92    & 100   & Y     & 175 \\
    45    & 46    & Y     & 175   & 94    & 100   & Y     & 175 \\
    46    & 47    & N     & 175   & 95    & 96    & Y     & 175 \\
    46    & 48    & Y     & 175   & 96    & 97    & Y     & 175 \\
    47    & 49    & Y     & 175   & 98    & 100   & Y     & 175 \\
    42    & 49    & Y     & 175   & 99    & 100   & Y     & 175 \\
    42    & 49    & Y     & 175   & 100   & 101   & Y     & 175 \\
    45    & 49    & Y     & 175   & 92    & 102   & Y     & 175 \\
    48    & 49    & Y     & 175   & 101   & 102   & N     & 175 \\
    49    & 50    & Y     & 175   & 100   & 103   & Y     & 500 \\
    49    & 51    & Y     & 175   & 100   & 104   & Y     & 175 \\
    51    & 52    & N     & 175   & 103   & 104   & Y     & 175 \\
    52    & 53    & Y     & 175   & 103   & 105   & Y     & 175 \\
    53    & 54    & Y     & 175   & 100   & 106   & Y     & 175 \\
    49    & 54    & Y     & 175   & 104   & 105   & Y     & 175 \\
    49    & 54    & Y     & 175   & 105   & 106   & Y     & 175 \\
    54    & 55    & Y     & 175   & 105   & 107   & Y     & 175 \\
    54    & 56    & Y     & 175   & 105   & 108   & N     & 175 \\
    55    & 56    & Y     & 175   & 106   & 107   & Y     & 175 \\
    56    & 57    & Y     & 175   & 108   & 109   & Y     & 175 \\
    50    & 57    & N     & 175   & 103   & 110   & Y     & 175 \\
    56    & 58    & Y     & 175   & 109   & 110   & Y     & 175 \\
    51    & 58    & Y     & 175   & 110   & 111   & Y     & 175 \\
    54    & 59    & Y     & 175   & 110   & 112   & Y     & 175 \\
    56    & 59    & Y     & 175   & 17    & 113   & Y     & 175 \\
    56    & 59    & Y     & 175   & 32    & 113   & Y     & 500 \\
    55    & 59    & Y     & 175   & 32    & 114   & N     & 175 \\
    59    & 60    & Y     & 175   & 27    & 115   & Y     & 175 \\
    59    & 61    & Y     & 175   & 114   & 115   & Y     & 175 \\
    60    & 61    & N     & 500   & 68    & 116   & Y     & 500 \\
    60    & 62    & Y     & 175   & 12    & 117   & Y     & 175 \\
    61    & 62    & Y     & 175   & 75    & 118   & Y     & 175 \\
    63    & 59    & Y     & 500   & 76    & 118   & Y     & 175 \\
\end{longtable}}

\begin{table}[!htbp]
  \centering
  \caption{Energy storage in the IEEE 118-Bus system}
  \small
    \begin{tabular}{lccr} \toprule
    Storage No. & Bus No. & Existing (Y/N) & Capacity (MW) \\
    \midrule
    S1     & 10     & N     & 20 \\
    S2     & 24     & N     & 20 \\
    S3     & 32     & N     & 20 \\
    S4     & 46     & N     & 20 \\
    S5     & 59     & N     & 20 \\
    S6     & 69     & N     & 20 \\
    S7     & 77     & N     & 20 \\
    S8     & 91     & N     & 20 \\
    S9     & 105     & N     & 20 \\
    S10     & 112    & N     & 20 \\
    \bottomrule
    \end{tabular}%
  \label{tab:118-bus-storage}%
\end{table}%

\end{document}